\PassOptionsToPackage{usenames,dvipsnames,svgnames,table}{xcolor}
\documentclass[acmtog]{acmart}
\usepackage{bm}
\copyrightyear{2026}
\acmYear{2026}
\setcopyright{cc}
\setcctype{by}
\acmConference[SA Conference Papers '26]{SIGGRAPH Asia 2026 Conference Papers}{December 01--04, 2026}{Kuala Lumpur, Malaysia}
\acmBooktitle{SIGGRAPH Asia 2026 Conference Papers (SA Conference Papers '26), December 01--04, 2026, Kuala Lumpur, Malaysia}
\acmDOI{10.1145/3829340.3842207}
\acmISBN{979-8-4007-2842-6/2026/12}

\acmSubmissionID{1380}

\usepackage{algorithm}
\usepackage{algorithmic}  %

\usepackage{booktabs}

\usepackage{caption}

\usepackage{array}   %
\usepackage{xcolor}  %

\definecolor{tabheader}{HTML}{FFFFE5}
\definecolor{tabgroupbase}{HTML}{EAF9FB}
\colorlet{tabgroup}{tabgroupbase!50}
\newcolumntype{C}[1]{>{\centering\arraybackslash\hspace{0pt}}p{#1}}

\graphicspath{{./figs_lowRes/}}

\usepackage{subcaption}

\usepackage{microtype}
\usepackage[normalem]{ulem}

\newif\ifsubmit
\submittrue
\ifsubmit
    \newcommand{\TODO}[1]{}
    \newcommand{\todo}[1]{}

    \newcommand{\yotam}[1]{}
    \newcommand{\yong}[1]{}

    \newcommand{\marginnote}[1]{}

    \newcommand{\revbegin}{}
    \newcommand{\revend}{}

    \newcommand{\deleted}[1]{}
    \newcommand{\deletedparagraph}[1]{}

    \newcommand{\revcaption}{}
\else
    \newcommand{\todo}[1]{{\color{red}[#1]}}
    \newcommand{\TODO}[1]{\textcolor{red}{\textbf{****** #1 ******}}}

    \newcommand{\sayit}[3]{{\small\protect\colorlet{col}{#2}\color{col}\colorbox{col!15}{\textsc{#1:}} #3}}

    \newcommand{\yotam}[1]{\sayit{Yotam}{purple}{#1}}
    \newcommand{\yong}[1]{\sayit{Yong}{green!50!black}{#1}}
    \newcommand{\marginnote}[1]{\marginpar{\colorbox{blue!10}{\begin{minipage}[t]{\linewidth}\tiny{#1}\end{minipage}}}}

    \newcommand{\revbegin}[1]{\color{blue}} %
    \newcommand{\revend}[1]{\color{black}} %

    \newcommand{\deleted}[1]{\textcolor{red}{\sout{#1}}}
    \newcommand{\deletedparagraph}[1]{\paragraph{\protect\sout{#1}}}

    \newcommand{\revcaption}{\captionsetup{labelfont={color=blue}}}
\fi

\makeatletter
\newcommand{\manuallabel}[2]{\def\@currentlabel{#2}\label{#1}}
\makeatother

\usepackage{float}
\usepackage{stfloats}

\usepackage{xr}
\makeatletter
\@for\@tempa:=-1,0,1,2,3,4,5\do{%
  \global\expandafter\let\csname r@tocindent\@tempa\endcsname\relax}%
\global\let\r@TotPages\relax  %
\makeatother

\begin{document}

\title{RBF Your SDF: Radial Basis Function Interpolation of Signed Distance Fields with Implied Tangent Points}

\author{Yong Cheng}
\email{ycheng27@gmu.edu}
\affiliation{%
  \institution{George Mason University}
  \country{USA}
}

\author{Yotam Gingold}
\email{ygingold@gmu.edu}
\affiliation{%
  \institution{George Mason University}
  \country{USA}
}
\begin{teaserfigure}
  \centering
  \includegraphics[width=\textwidth]{teaser/teaser.pdf}
  \makebox[\textwidth][r]{\includegraphics[width=0.985\textwidth]{teaser/denker_k.pdf}}
  \caption{
  We marry scattered data interpolation with the tangent sphere observation \cite{sellan2023rfts,batty_sharp_2011,kobbelt_feature_2001} that every Signed Distance Field (SDF) sample implies a point on a sphere tangent to the surface.
  We propose an iterative procedure to find tangent points consistent with a Radial Basis Function (RBF) interpolating both SDF and tangent point samples.
    Our algorithm is practical at higher resolutions via a partition-of-unity (PU) formulation
    and produces lower error at all tested grid sizes.}
  \label{fig:teaser}
\end{teaserfigure}

\begin{abstract}
Signed distance fields (SDFs) are a popular implicit
representation of geometry. Converting a discrete set of
SDF samples into an explicit surface is a fundamental problem in
geometry processing. Traditional reconstruction methods such as marching cubes and dual contouring ignore the geometric information carried by samples far from the surface.
Recently, \citet{sellan2023rfts} and several follow-up works leveraged the tangent-sphere structure of SDFs; every sample implies a point on a sphere tangent to the surface.
However, these approaches extract the zero-level set via
surface reconstruction, which considers only points and normals
on the surface and ignores the remaining samples.
We propose an approach that marries the tangent-sphere observation with radial basis function
interpolation of all data, the implied surface points
and the original data.
By detecting spheres with extremely constrained tangent points, a configuration geometrically forced at
sharp surface features, we identify and preserve surface corners
that surface reconstruction-based methods systematically round. A
partition-of-unity decomposition allows our method to scale
efficiently to large grid resolutions. Our
reconstructions improve both Chamfer and Hausdorff accuracy
at every tested resolution.
\end{abstract}

\maketitle

\section{Introduction}

Reconstructing a surface from a signed distance function (SDF) sampled on a grid or point set is a fundamental problem in geometry processing, with applications in shape modeling, physical simulation, and neural implicit representations.
Widely used approaches to zero level-set extraction (e.g., Marching Cubes~\cite{lorensen1987marching}, Marching Tetrahedra~\cite{doi1991marching}, and Dual Contouring~\cite{ju2002dual}) rely on the local information within cells of a tessellation.
Recently, \citet{sellan2023rfts} observed that every SDF sample $(\mathbf{x}_i, d_i)$ implicitly defines a sphere of radius $|d_i|$ centered at $\mathbf{x}_i$ that is tangent to the surface (and whose interior lies entirely outside or inside the surface).\footnote{They credit \citet{batty_sharp_2011} and \citet{kobbelt_feature_2001} for this observation.}
By finding a point on every tangent sphere (and its implied normal),
follow-up work \cite{sellan2024rfta,kohlbrenner2025empty}
achieved state-of-the-art performance
by leveraging the well-studied
problem of surface reconstruction \cite{kazhdan2013screened}
from a set of surface points and normals.
However, by relying on surface reconstruction,
we claim that these tangent-sphere approaches
discard useful information from the original samples
and become too sensitive to the quality of their chosen points and normals. Moreover, surface reconstruction based on points and normals struggles with sharp feature points where no one normal describes the local geometry.

At the same time, scattered data interpolation provides a general solution to the problem of reconstructing a continuous function from discrete samples (see \S\ref{sec:related}). Radial basis function (RBF) interpolation, is a classical choice for reconstructing smooth scalar fields from scattered samples. However, although the RBF interpolant agrees with the input SDF at every sample, it does not recover the correct zero level-set. Because no sample lies exactly on the surface, the reconstructed zero level-set is shifted away from the true surface (Figure~\ref{fig:eiffel}).

Our work marries scattered data interpolation with the tangent sphere observation.
We construct an RBF that interpolates each SDF sample
$(\mathbf{x}_i, d_i)$
along with a paired zero-valued sample on its tangent sphere
$(\mathbf{y}_i, 0)$, where $\|\mathbf{x}_i-\mathbf{y}_i\|=|d_i|$.
We propose an iterative procedure to find tangent points consistent with the RBF interpolant.
We provide an explicit construction for the region of allowable tangent points on each sample sphere, a formula for its area, and an efficient approximation.
We use the cubic kernel, which approximately preserves the Eikonal constraint.
To make interpolatory RBF reconstruction practical at high resolutions, we adopt a partition-of-unity (PU) 
formulation~\cite{franke_smooth_1982,ohtake2003mpu} that decomposes the domain into overlapping local patches, 
each solved independently and blended with smooth weights, while remaining strictly interpolatory within
each patch.
We evaluate our method on a $300$-mesh subset of
Thingi10K~\cite{Zhou2016Thingi10K} at grid resolutions from $6^3$ to $100^3$. Our method improves both Chamfer and Hausdorff accuracy at every tested resolution.

\section{Related Work}
\label{sec:related}

\subsection{Surface extraction from SDF samples}
Classical, widely-used approaches to explicit surface extraction from sampled implicit functions
are Marching Cubes~\cite{lorensen1987marching},
Marching Tetrahedra~\cite{doi1991marching,treece1999regularised},
and the dual variant known as Dual Contouring~\cite{ju2002dual,hwang2024odc}.
(See \citet{hwang2024odc} for a recent such work and \citet{de_araujo_survey_2015} for a survey that includes alternative approaches.)
These methods assume a spatial tessellation and examine the
function values in a single cell at a time.
They ignore the global
geometric information (tangent-sphere property) carried by samples further from the surface.
Learning-based variants such as Neural Marching
Cubes~\cite{chen2021nmc} and Neural Dual
Contouring~\cite{chen2022ndc} mitigate this by predicting vertex
placements from larger local windows and data-driven assumption, but they are more expensive, still only consider local windows, and operate on fixed grids. (They may implicitly make use of the tangent-sphere property via the larger windows.)
Our approach first reconstructs a higher quality continuous function (via tangent-spheres and scattered data interpolation) and then relies on Dual Contouring to extract the final mesh.

\citet{sellan2023rfts} credit \citet{batty_sharp_2011} and \citet{kobbelt_feature_2001}
with the observation that every SDF sample implies a sphere tangent to (and otherwise disjoint from) the surface.
\citet{sellan2023rfts} exploit this observation in
\emph{Reach for the Spheres} (RFTS), an energy-driven mesh flow
that improves reconstruction quality but is restricted to the
topology of the initial mesh. Their follow-up
\emph{Reach for the Arcs}
(RFTA)~\cite{sellan2024rfta} samples tangent points on the
intersection-free portion of each sphere and feeds the resulting
oriented point cloud to screened
Poisson Surface Reconstruction (sPSR)~\cite{kazhdan2013screened}, removing
the topological restriction.
The RFTA algorithm optimizes for a set of tangent points that agree with their sPSR.
\citet{kohlbrenner2025empty} pursue the same overall pipeline
deterministically. They use Lie sphere geometry to construct the
maximal empty spheres (MES) in the SDF data, take the resulting contact
points as oriented surface samples, and likewise feed them to
screened Poisson Surface Reconstruction, avoiding the iterative point
optimization of RFTA.
More recently, the same authors approximate each sample's gradient
from a triangulation of the sample
locations~\cite{kohlbrenner2026contouring}, take
$\mathbf{x}_i - d_i\,\mathbf{g}_i$ as a surface point, and discard
candidates falling inside another sample's sphere. The gradient estimate is
purely local. It degrades near the medial axis, where
averaging across the crease yields
an invalid direction likely to be filtered away.
We instead
evaluate our interpolant at the projected location and keep the
direction whose projection reaches farthest into the opposite
half-space
(Sec.~\ref{sec:iterative-projection}); this selects one of the
several contact directions rather than an average of them.
\citet{ren2024mcgrids}'s work on adaptive
refinement formulates point insertion as a
Monte Carlo sampling problem but requires SDF queries at arbitrary
locations.
These works obtain results far superior to classical approaches (e.g., Marching Cubes and Dual Contouring).
We are inspired by their approaches.
However, instead of relying on surface reconstruction, which relies solely on surface points (and normals),
we use both the tangent points (distance $=0$) and original points (distance $\neq 0$) in a scattered data interpolation algorithm to obtain a higher-quality continuous function.

\subsection{Power-diagram-based SDF contouring}
Power diagrams have long been used for surface
reconstruction. The power crust of \citet{amenta2001powercrust}
approximates the medial axis transform of an object by \emph{polar
balls} derived from the Voronoi diagram of a dense sample of points
\emph{on} its surface, labels the cells of the balls' power diagram as
inside or outside, and outputs the power-diagram faces separating the
two as the reconstructed surface.
Two recent works~\cite{kohlbrenner2025power, wang2025power} apply
the same machinery in the SDF setting, where the balls come directly
from the samples. They build on
the tangent-sphere observation %
by constructing power diagrams of weighted SDF samples, where each
sample is weighted by its squared signed distance. \citet{kohlbrenner2025power} extract the surface as either a
primal marching-tetrahedra contour on the regular triangulation or
its dual power contour. They additionally propose a refinement scheme
(RCR) that incrementally inserts new sites at locations chosen from
the dual power vertices; this scheme requires the ability to query
the SDF at arbitrary new locations and is therefore inapplicable
when only a fixed set of samples is available---the setting we and
RFTS/RFTA address. On fixed samples alone, the authors report that
RFTA still outperforms their primal and dual variants.
\citet{wang2025power} likewise construct a power diagram,
augmented with surface projection points
$\mathbf{p}_i - \phi(\mathbf{p}_i)\nabla\phi(\mathbf{p}_i)$ derived
from per-sample gradients; their method thus assumes both SDF values
\emph{and} gradients as input, in contrast to the SDF-only setting of
RFTS, RFTA, and our method.

\subsection{Scattered data interpolation and radial basis functions}
Scattered data interpolation seeks a function that passes through
prescribed values, and possibly derivatives, at arbitrary point
locations. Radial basis functions (RBFs) are a classical and
well-studied tool for this task \cite{duchon1977splines, wendland2004scattered,
anjyo2014scattered}. \citet{carr2001reconstruction} and
\citet{turk2002modelling} pioneered the use of RBF
interpolation to recover surfaces from oriented point clouds (e.g., surface points) by
fitting an implicit field to point positions augmented with
off-surface constraints. To handle inputs without normals,
\citet{huang2019vipss} (VIPSS) and the follow-up
\emph{NN-VIPSS}~\cite{xia2025nnvipss} fit a Hermite RBF~\cite{macedo2011hermite}
while jointly optimizing the unknown sample normals.
We apply RBF interpolation to the input SDF samples, which in general
do not lie on the surface, together with tangent-sphere points.
Because the values are not all zero, we do not need normal information.
Similar to NN-VIPSS, we also employ a partition of unity scheme for acceleration.
\citet{drake2022cfpu} proposed curl-free RBF kernels in a partition-of-unity formulation
to reconstruct surfaces from oriented point clouds at scale.
We experimented with this approach, but obtained lower-quality surfaces.

\section{Background}
\label{sec:preliminaries}

\paragraph{Problem setup.}
Let $\Omega \subset \mathbb{R}^3$ be an unknown closed surface
that we wish to reconstruct, and let $D : \mathbb{R}^3 \to
\mathbb{R}$ denote its signed distance function, with the
convention that $D$ is negative in the interior and positive
on the exterior, so that
$\Omega = \{\mathbf{x} \in \mathbb{R}^3 : D(\mathbf{x}) = 0\}$.
We are given $n$ samples
$\{(\mathbf{x}_i, d_i)\}_{i=1}^n$ with $d_i = D(\mathbf{x}_i)$,
either on a regular grid or scattered. Our goal is to recover a
continuous function $\tilde D : \mathbb{R}^3 \to \mathbb{R}$ that
\emph{interpolates} the samples, $\tilde D(\mathbf{x}_i) = d_i$
for all $i$, and whose zero level set $\{\tilde D = 0\}$ closely
approximates $\Omega$.

\subsection{Tangent-sphere structure of SDF samples}
\label{sec:tangent}

Each sample carries more information than its scalar value alone.
As observed by prior work \citet{batty_sharp_2011,sellan2023rfts,kobbelt_feature_2001}, the unsigned
value $|d_i|$ defines a sphere
$S_i := \{\mathbf{x} \in \mathbb{R}^3 :
  \|\mathbf{x} - \mathbf{x}_i\| = |d_i|\}$
whose enclosed ball lies entirely on one side of $\Omega$ (interior
if $d_i < 0$, exterior otherwise), and which is tangent to $\Omega$
at the closest surface point
\begin{equation}
  \label{eq:tangent}
  \mathbf{a}_i \;=\; \mathbf{x}_i - d_i\,\nabla D(\mathbf{x}_i)
  \;\in\; S_i,
  \qquad
  D(\mathbf{a}_i) = 0.
\end{equation}
Each input sample thus implies an additional zero-valued
constraint located on $S_i$.
The position $\mathbf{a}_i$, however,
is not directly available, since it depends on the unknown
gradient $\nabla D(\mathbf{x}_i)$; recovering an estimate of
$\mathbf{a}_i$ is the core algorithmic step described in
Section~\ref{sec:iterative-projection}.
Every tangent sphere surface has at least one point not contained in the interior of other spheres;
we call the set of such points the \emph{exposed region}.\footnote{\citet{sellan2024rfta} call this the
\emph{feasibility arc}. We avoid this term to prevent confusion with the arcs which bound the exposed region.}

\subsection{Radial basis function interpolation}
\label{sec:rbf-prelim}

For data $\{(\mathbf{x}_i, d_i)\}_{i=1}^n \subset
\mathbb{R}^3 \times \mathbb{R}$, a radial basis function (RBF)
interpolant takes the form~\cite{wendland2004scattered}
\begin{equation}
  \label{eq:rbf}
  f(\mathbf{x}) = \sum_{i=1}^{n} \alpha_i\,
    \phi(\|\mathbf{x} - \mathbf{x}_i\|)
    + p(\mathbf{x}),
\end{equation}
where $\phi : \mathbb{R}_{\geq 0} \to \mathbb{R}$ is a fixed
radial kernel and $p$ a low-order polynomial. We use the
\emph{cubic polyharmonic} kernel $\phi(r) = r^3$ together with
a degree-$1$ polynomial $p(\mathbf{x}) = \mathbf{c}^\top
\mathbf{x} + c_3$. The coefficients $\boldsymbol\alpha \in \mathbb{R}^n$
and $(\mathbf{c}, c_3) \in \mathbb{R}^4$ are determined by the
$n$ interpolation conditions $f(\mathbf{x}_i) = d_i$ together with
the four moment conditions $\sum_i \alpha_i \mathbf{x}_i = \mathbf 0$
and $\sum_i \alpha_i = 0$, yielding the symmetric
saddle-point system
\begin{equation}
  \label{eq:rbf-system}
  \begin{pmatrix} A & P \\ P^\top & 0 \end{pmatrix}
  \begin{pmatrix} \boldsymbol\alpha \\ \mathbf{c}_{0:3} \end{pmatrix}
  =
  \begin{pmatrix} \mathbf{d} \\ \mathbf 0 \end{pmatrix},
  \qquad A_{ij} = \|\mathbf{x}_i - \mathbf{x}_j\|^3,
\end{equation}
with $P \in \mathbb{R}^{n\times 4}$ the polynomial design matrix 
whose $i$-th row is $(\mathbf{x}_i^\top,1)$,
and $\mathbf{c}_{0:3} = (c_0, c_1, c_2, c_3)^\top$. The cubic
kernel is conditionally positive definite of order $2$
(\cite{wendland2004scattered}, Corollary 8.18), so a
linear polynomial augment is the minimum required for unique
solvability.

We selected the cubic (3D triharmonic) kernel $\phi(r) = r^3$,
after empirically comparing it against $r$, $r^2 \log r$, the
Gaussian kernel, and Wendland's compactly supported $C^2$ kernel.
The cubic kernel best preserves the unit-norm gradient property
of a true SDF between data points and in the far field (Appendix Figure~\ref{fig:kernels}).
This can be explained in 1D, where evaluating the cubic kernel at $x$ ``to the right of'' the span of data $x_i$ lets us 
expand the cubic kernel without absolute value, i.e.,
\[
\sum_i \alpha_i\left(x-x_i\right)^3=x^3 \sum_i \alpha_i-3 x^2 \sum_i \alpha_i x_i+3 x \sum_i \alpha_i x_i^2-\sum_i \alpha_i x_i^3.
\]
The zero-moment conditions eliminate the cubic and quadratic terms 
$x^3 \sum \alpha_i - 3x^2 \sum \alpha_i x_i$, leaving behind a linear function.
\label{sec:cubic-kernel-extrapolation}

We do not regularize the system.
Replacing the
kernel block $A$ in \eqref{eq:rbf-system} with $A+\lambda I$
(Tikhonov) does help on the occasional ill-conditioned input in \citet{drake2022cfpu}, but
on most inputs it smooths out the details exact interpolation aims
to preserve. We instead eliminate the near-coincident constraints
that motivate it~\cite{wendland2004scattered} upstream by
de-duplication (\S\ref{sec:deduplication}).

\begin{figure*}[t]
    \centering
    \includegraphics[width=.9\linewidth]{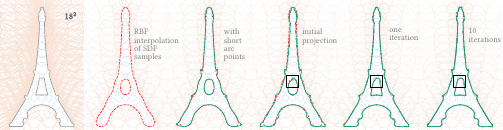}
    \caption{Progressive 2D reconstruction of the \textsc{eiffel}
    outline from $18^2$ SDF samples. From left to right:
    (i)~ground truth;
    (ii)~the zero level set of the interpolant of the SDF samples alone, which
    floats off the true surface and rounds off every sharp feature;
    (iii)~after adding short-arc tangent points
    (Sec.~\ref{sec:short-arc}), which restore the corners;
    (iv)~after one outer iteration, with samples whose spheres have
    no intersecting neighbor deferred from projection;
    (v)~the next iteration projects every remaining sample, but
    those failing the feasibility test are rejected;
    (vi)~after $10$ iterations, additional projections become
    feasible and the level set has converged to the true surface.
    Insets in the last 3 panels show how iterations smooth the
    level set near tangencies. Sample circles are faded by radius for clarity in
    panels (ii)--(vi). Tangent projections are drawn as light-green
    dots. For comparison, panels (iii) and (iv)
    overlay the level set (green) on (ii) (red dashed); (v) and (vi)
    overlay it on the ground truth (gray).
    }
    \label{fig:eiffel}
\end{figure*}

\begin{figure}[t]
    \centering
    \includegraphics[width=\linewidth]{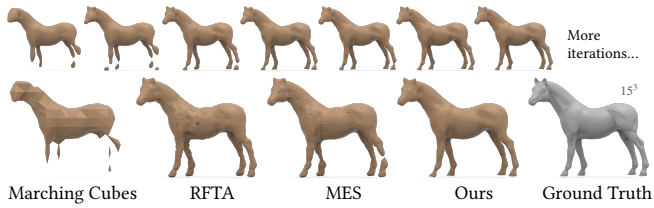}
    \caption{Reconstruction of the \textsc{horse} mesh at $15^3$
    sampling. \emph{Top row:} Our zero level set across iterations,
    starting from the interpolant of the SDF samples alone on the left and refined
    by successive iterative-projection steps to the right.
    \emph{Bottom row:} Output versus ground truth. Our output mesh
    is closer to the ground truth geometrically and topologically
    compared to alternatives.
    }
    \label{fig:horse}
\end{figure}

\section{Algorithm}
\label{sec:algorithm}
Given SDF samples $\{(\mathbf{x}_i, d_i)\}_{i=1}^n$, our algorithm (Figures \ref{fig:eiffel} and \ref{fig:horse})
recovers a smooth (except at sharp features) and accurate interpolatory $\tilde D$,
followed by mesh extraction.
The goal of our algorithm is to find a paired tangent point $\mathbf{y}_i$ for each sample $(\mathbf{x}_i, d_i)$.
The core of our algorithm is straightforward:
\begin{algorithmic}
\WHILE{not converged}
\STATE $\mathbf{y}_i \gets$ Estimate tangent points for samples $\mathbf{x}_i$
\STATE $\tilde D \gets \mathrm{RBF}( \{ ( \mathbf{x}_i, d_i ) \} \cup \{ \mathbf{y}_i, 0 \})$
\ENDWHILE
\RETURN $\textsc{IsosurfaceExtract}(\tilde D, 0)$
\end{algorithmic}
See Appendix~\ref{supp:pseudocode} for in-depth pseudocode and a description of our partition of unity scheme.

The first iteration extracts tangent points for spheres whose exposed regions have collapsed to (nearly) a point (Section~\ref{sec:short-arc}).
We then alternate between computing an
interpolatory partition-of-unity RBF $\tilde D$ over the samples $( \mathbf{x}_i, d_i )$ and current tangent points $\mathbf{y}_i$,
and updating each sample sphere's tangent direction by an $S^2$
search using the
current $\tilde D$ as a guide
(Section~\ref{sec:iterative-projection}).
Both steps of the algorithm are parallelizable, the first step since tangent points can be estimated independently, and the second step because of partition of unity.
Finally, because $\tilde D$ is defined everywhere, one can sample it
on a new grid at any resolution and apply any standard contouring
algorithm to extract the mesh. We extract the
surface mesh from the zero level set of $\tilde D$
using Dual Contouring \cite{ju2002dual} at
a per-axis extraction resolution of
$\min\bigl(\max\bigl(\lceil 4(\sqrt[3]{n}-1)+1\rceil,\,64\bigr),\,512\bigr)$.
The bounds keep the output usable for very sparse inputs and cap the
extraction cost for very dense ones.

\begin{figure}
    \centering
    \begin{subfigure}[b]{0.36\columnwidth}
        \centering

        \includegraphics[width=.8\linewidth]{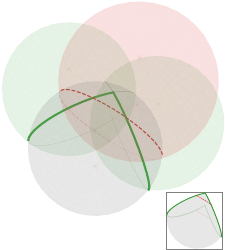}
        \caption{}
        \label{fig:cap-update-step1}
    \end{subfigure}\hfill
    \begin{subfigure}[b]{0.36\columnwidth}
        \centering
        \includegraphics[width=.8\linewidth]{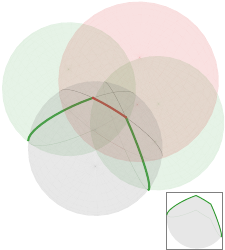}
        \caption{}
        \label{fig:cap-update-step2}
    \end{subfigure}\hfill
    \begin{subfigure}[b]{0.24\columnwidth}
        \centering
        \includegraphics[width=\linewidth]{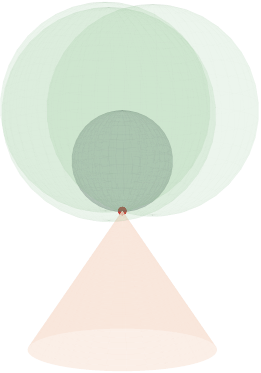}
        \caption{}
        \label{fig:cap-update-corner}
    \end{subfigure}
    \vspace{-1em}
    \caption{Per-sphere incremental update of the exposed region
    boundary on a sample sphere (gray).
    (\subref{fig:cap-update-step1}) and (\subref{fig:cap-update-step2})
    show the same update in 2D (top) and 3D (bottom).
    (\subref{fig:cap-update-step1})~Existing caps and their currently
    exposed arcs (green); the incoming cap is shown as a dashed red
    circle.
    The exposed region is shown inset.
    (\subref{fig:cap-update-step2})~After the new cap is committed,
    existing green arcs are clipped where they enter the new cap and
    the new cap's surviving arcs (red) are added to the boundary.
    The exposed region (inset) shrinks accordingly.
    (\subref{fig:cap-update-corner})~At a sharp corner, neighboring
    sample spheres bracket the arcs from many directions and the
    exposed region collapses to the corner point (red).
    }
    \label{fig:cap-update}
\end{figure}

\subsection{Geometrically Determined Tangent Points}
\label{sec:short-arc}
\begin{figure}[t]
    \centering
    \begin{subfigure}[b]{0.42\columnwidth}
        \centering
        \includegraphics[width=\linewidth]{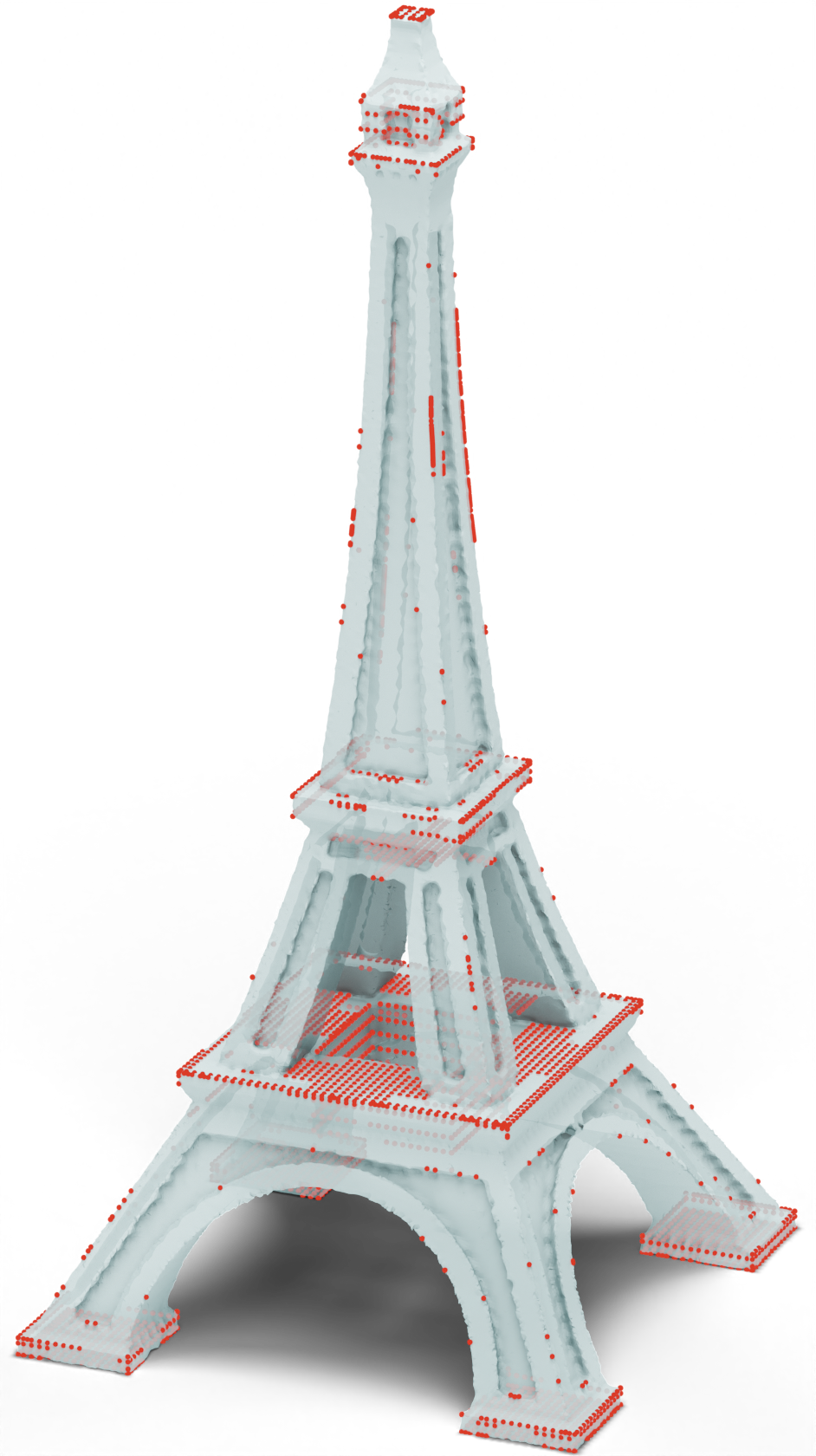}
        \caption{}
        \label{fig:short-arc-fast}
    \end{subfigure}
    \hfill
    \begin{minipage}[b]{0.55\columnwidth}
        \begin{subfigure}[b]{\linewidth}
            \centering
            \includegraphics[width=\linewidth]{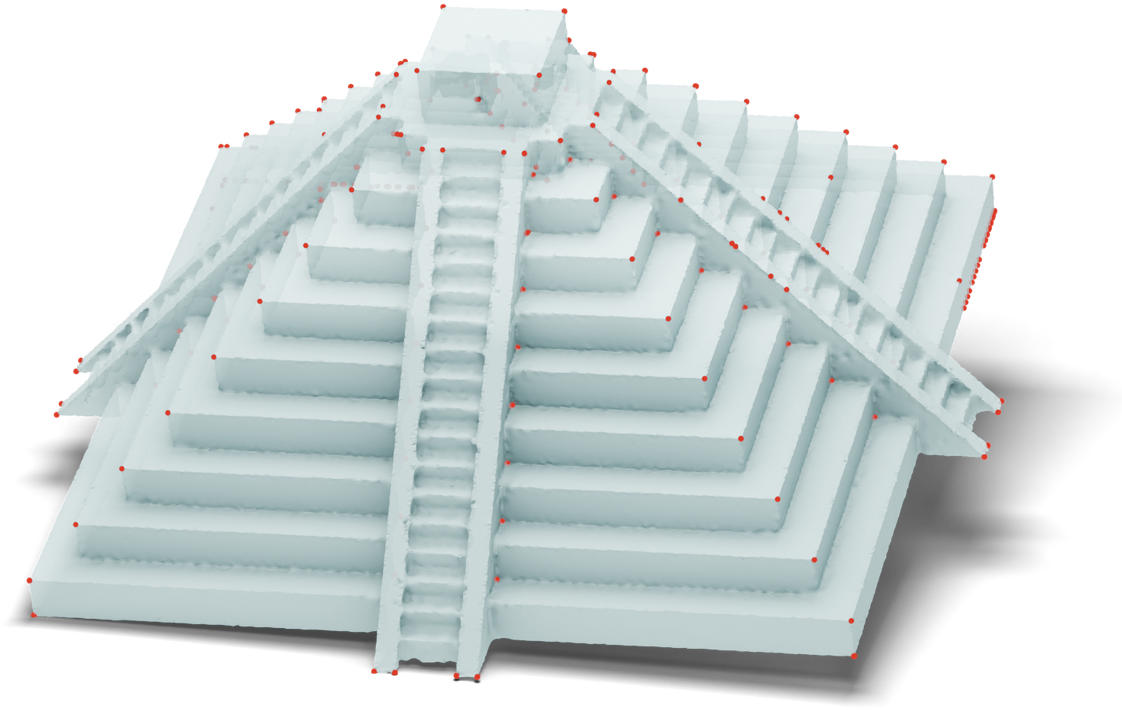}
            \caption{}
            \label{fig:pyramid-rot}
        \end{subfigure}\\[1ex]
        \begin{subfigure}[b]{\linewidth}
            \centering
            \includegraphics[width=\linewidth]{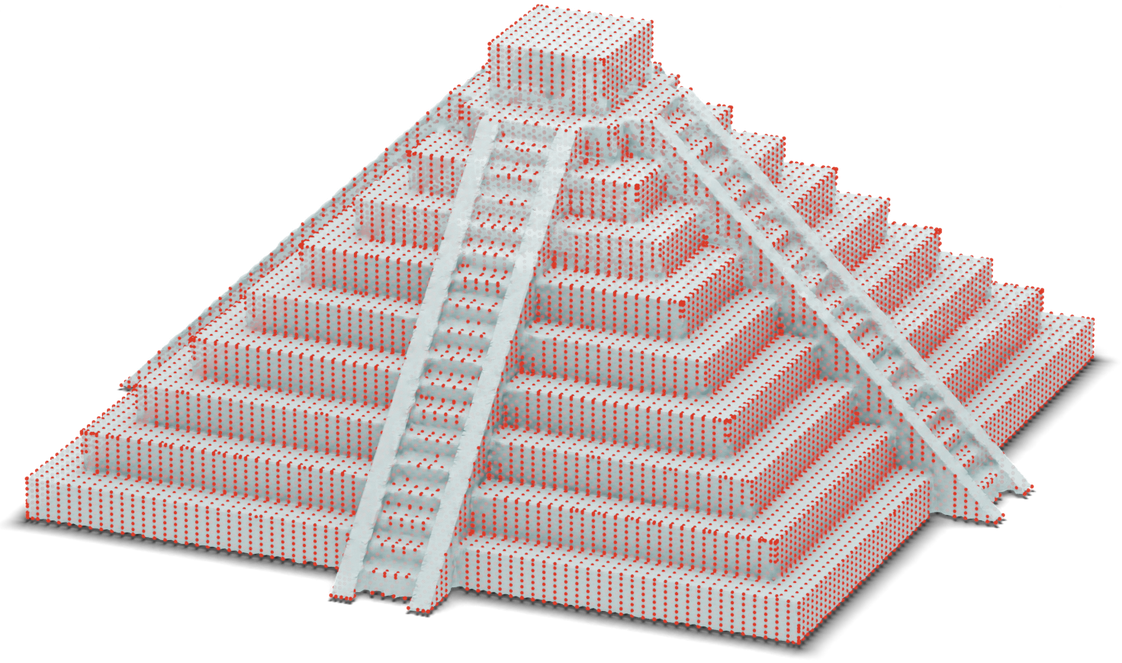}
            \caption{}
            \label{fig:pyramid}
        \end{subfigure}
    \end{minipage}
    \vspace{-1em}
    \caption{Small exposed regions' tangent points (red dots) at $n=100^3$.
    A reconstructed (\subref{fig:short-arc-fast}) \textsc{Eiffel} and (\subref{fig:pyramid-rot}) \textsc{Pyramid} surface.
    On an axis-aligned \textsc{Pyramid} (\subref{fig:pyramid}), collinear spheres project to a common point on flat surfaces as well.
    }
    \label{fig:short-arc-examples}
    \vspace{-1em}
\end{figure}

In two cases, the tangent point is completely determined by geometry alone.
(A) At a sharp surface feature, a sample sphere can be squeezed by many
neighbors until its exposed region collapses to (nearly) a single point (Figure~\ref{fig:cap-update-corner}).
This happens most reliably at corners (where the sphere is
bracketed from all directions) but also along sharp edges when the
sampling along the edge is dense enough to bracket a smaller sphere.
(B) When two spheres are \emph{inner tangent}, the smaller one is strictly inside 
the larger one except at the tangent point while $\|\mathbf{x}_i-\mathbf{x}_j\| = |d_i-d_j|$ (Figure~\ref{fig:pyramid}).
Both of these situations can be detected by computing the size of exposed regions.
We emit such completely determined tangent points as zero-valued interpolation constraints,
bypassing the iterative projection step
(Section~\ref{sec:iterative-projection}) required for general sample
spheres.

\paragraph{Small exposed regions}
To compute the size of an exposed region, we first compute the set of arcs on $S_i$ that bound it. Given such arcs, we can precisely
compute the area of the exposed region via Gauss-Bonnet,%
\footnote{Gauss-Bonnet states $\int_M K \, dA + \int_{\partial M} \kappa_g \, ds = 2\pi \chi(M)$. Because $M$ is a subset of a sphere with radius $r$, $\int_M K \, dA = \frac{A}{r^2}$, where $A$ is the unknown area of $M$, and $\chi(M) = C - L$, where $C$ are the number of exposed region connected components and $L$ the number of holes in them. Rearranging terms, the formula is $A = r^2( 2\pi (C-L) - \int_{\partial M} \kappa_g \, ds )$. The integrated geodesic curvature along a boundary loop of piecewise circular arcs on the sphere is $\sum \frac{h_i}{r} \phi_i + \sum \theta_{i \rightarrow j}$, where $h_i$ is the distance of circular arc $i$'s plane from the sphere center, $\phi_i$ the arc's radians (i.e., $2\pi$ for a complete circular arc), and $\theta_{i\rightarrow j}$ is the exterior turning angle (signed angle between tangents on the sphere, positive when turning towards the interior) where arc $i$ meets arc $j$.}
though we observe that perimeter collapse is sufficient to find tiny regions,
requires no connectivity of arcs, and is perhaps even preferable since slivers may have small area but leave substantial ambiguity as to the location of the tangent point.
Exposed regions whose total arc length falls below
$\epsilon_{\mathrm{degen}}$ are treated as collapsed to a point.

Given a sample sphere $S_i$ with center $\mathbf{c}_i$ and radius $r_i$,
each intersecting neighbor $S_j$ (let $D_{ij} = \lVert\mathbf{c}_j-\mathbf{c}_i\rVert$)
carves a spherical cap $C_j$ on $S_i$, bounded by a circle that lies in the
cutting plane $\mathbf{n}_{ij}\!\cdot\mathbf{x} = d_{ij}$ with
\begin{equation}
    \mathbf{n}_{ij} = \frac{\mathbf{c}_j - \mathbf{c}_i}{D_{ij}},
    \qquad
    h_{ij} = \frac{D_{ij}^2 + r_i^2 - r_j^2}{2\,D_{ij}},
    \qquad
    d_{ij} = \mathbf{n}_{ij}\!\cdot\mathbf{c}_i + h_{ij}.
    \label{eq:cap-params}
\end{equation}
The circle bounding $C_j$ has radius
$\rho_j$$= \sqrt{r_i^2 - h_{ij}^2}$.

We process neighbors of $S_i$ one at a time, maintaining the running list of
boundary arcs $\{a_\ell\}_\ell$, each stored as an angular interval
$[t_\ell^{\mathrm{start}}, t_\ell^{\mathrm{end}})$ on the circle of its
host cap $C_{c(\ell)}$. When a new cap $C_{\mathrm{new}}$ is added,
the update is bidirectional (Figure~\ref{fig:cap-update}):
\begin{enumerate}
    \item \emph{Clip existing arcs by $C_{\mathrm{new}}$.} For every arc
    $a_\ell$, intersect its host circle with
    $C_{\mathrm{new}}$'s cutting plane, then drop the angular portion of
    $a_\ell$ that lies inside $C_{\mathrm{new}}$.
    \item \emph{Compute $C_{\mathrm{new}}$'s own arcs against the preceding caps.}
    Starting from the full circle $[0, 2\pi]$ on $C_{\mathrm{new}}$, intersect
    with the plane of each previously processed cap.
    The surviving intervals are
    $C_{\mathrm{new}}$'s contribution to the boundary.
\end{enumerate}

Pseudocode for this process appears in Appendix~\ref{supp:pseudocode}.

\paragraph{Tangent points at near degeneracies.}
The interval intersection in step~(2) tolerates a small ``negative'' arc.
A preceding cap's plane may cut the $C_{\mathrm{new}}$ circle just outside its arc interval.
If this would create a short negative interval ($<\epsilon_{\mathrm{tan}} = 10^{-4}\,\text{rad}$),
we emit a
\emph{zero-length} arc at the midpoint of the negative arc instead of discarding it.

If, after processing all neighbors, the total arc length on $S_i$ falls
below a small threshold $\epsilon_{\mathrm{degen}}$,
\begin{equation}
    \sum_\ell \rho_{c(\ell)}\bigl(t_\ell^{\mathrm{end}} - t_\ell^{\mathrm{start}}\bigr) \;<\; \epsilon_{\mathrm{degen}},
\end{equation}
the exposed region has shrunk to (essentially) a point (Figures~\ref{fig:cap-update-corner} and \ref{fig:short-arc-examples}). We emit each arc's
midpoint directly as a tangent-point candidate.
A sphere may have several disjoint small exposed regions. We keep the arc
midpoint with the smallest $|\tilde D_0|$, where $\tilde D_0$ is the interpolant
fit to the SDF samples alone, and only if
$|\tilde D_0| < \epsilon_{\mathrm{val}} = 0.1$. If any candidate is kept, it
becomes a committed tangent point and its sphere skips the iterative projection step (Sec.~\ref{sec:iterative-projection}).

We use $\epsilon_{\mathrm{degen}} = 10^{-5}$ for meshes normalized to a
unit bounding box. Experimentally, the algorithm is insensitive to this
choice across $\epsilon_{\mathrm{degen}} \in [10^{-8}, 10^{-4}]$ (Appendix~\ref{supp:degen-tol}).

\paragraph{Computational Complexity.}
For a sample sphere with $k$ intersecting neighbors, clipping each cap
circle against the remaining $k-1$ caps costs $O(k)$, so the per-sphere
work is $O(k^2)$.
Aggregating across $n$ samples gives $O(n\,k^2)$.
Naively, $k$ scales linearly with $n$ as the sampling density grows.
Refining the sampling inserts new spheres between any two that already 
overlap, and those new spheres must overlap both, so each sphere's intersection
count grows \emph{linearly} with the total sample count $n$.
Supplementary Figure~\ref{fig:phantom-and-k}(\subref{fig:k-vs-n}) shows both the median and the
maximum neighbor count scaling linearly with $n$.
In the worst case, $k=n$ and the complexity is $O(n^3)$.

We avoid this scaling,
as only an $O(1)$ (non-pathological) subset of neighbors actually contribute arcs to its exposed region, reducing the total cost to $O(n)$ plus the $O(n \log n)$ (non-pathological) cost of finding appropriate neighbors via a \emph{regular triangulation} (RT).
Two spheres share exposed region boundary arcs only if their samples are neighbors in the RT of all samples~\cite{edelsbrunner1995union,kohlbrenner2025power,
kohlbrenner2026contouring}.
A sample with a small exposed region, however, is
numerically fragile
and may be omitted from the RT.
In that case, its appropriate neighbor set
is the vertices of the RT simplex containing
it.

\subsection{Iterative Projection}
\label{sec:iterative-projection}

\paragraph{Direction refinement.}
Tangent spheres without small exposed regions determine their tangent points iteratively.
A natural approach to locate where each sample's tangent sphere
touches the surface is to extract
a mesh from the current SDF and perform closest-point queries against it~\cite{sellan2024rfta}.
This, however, ties the projection accuracy to mesh resolution. At low
resolution, many projections collapse onto the same vertex, while at high
resolution the cost is dominated by isosurface extraction. We instead refine
the tangent direction directly on $S^2$ in a mesh-free manner.

Specifically, for each sample $\mathbf{x}_i$ with $d_i = \tilde D(\mathbf{x}_i)$,
we seek the unit direction $\mathbf{g}_i$ for which the projected point
$\mathbf{y}_i = \mathbf{x}_i - d_i\,\mathbf{g}_i$ is tangent to the surface, i.e.,
$\mathbf{g}_i$ aligns with the outward unit normal
$\hat{\mathbf{n}}(\mathbf{y}) := \nabla\tilde D(\mathbf{y})/\lVert\nabla\tilde D(\mathbf{y})\rVert$
at $\mathbf{y}_i$.
In other words, among all points $\mathbf{y}$ on the sphere
$\lVert\mathbf{y}-\mathbf{x}\rVert = |d|$, we seek the one that reaches
farthest past the surface into the region of opposite sign, so we minimize
\begin{equation}
    f(\mathbf{g}) \;=\; \tilde D(\mathbf{x} - d\,\mathbf{g})\,/\,d
\end{equation}
over $\mathbf{g}\in S^2$.
Dividing by the signed distance $d$ both
orients the objective for samples of either sign and cancels the chain
rule, so the Euclidean gradient is exactly
$\nabla f = -\nabla\tilde D(\mathbf{y})$.
At a stationary direction $\mathbf{g}^{\!*}$, the tangential part of
$\nabla\tilde D(\mathbf{y})$ vanishes, i.e.,
$\mathbf{g}^{\!*} \parallel \nabla\tilde D(\mathbf{y}^{\!*})$ and
$\mathbf{g}^{\!*}$ agrees with $\hat{\mathbf{n}}$ as required.

We minimize $f$ with BFGS~\cite{nocedal2006numerical},
normalizing after each step to project onto the sphere.
We terminate once the tangential gradient
falls below $10^{-4}$ ($\approx 0.006^{\circ}$)
or after five evaluations, since the next outer iteration
re-solves the samples against an updated $\tilde D$ anyway.
Tangential steps are capped at $0.2$ initially and at $1$ thereafter,
rotating $\mathbf{g}$ by at most $11^{\circ}$ and $45^{\circ}$.

\paragraph{Initialization.}
Small-exposed-region samples handled by the short-arc step
(Section~\ref{sec:short-arc}) take no part in the iterative
refinement at any iteration. For the remaining samples, every
iteration warm-starts from the previous iteration's converged
$\mathbf{g}$. The very first $\mathbf{g}^{(0)}$ comes from a
Fibonacci-lattice search: we sample $64$ near-uniform directions
$\mathbf{u}_j$ on $S^2$ and pick the one minimizing
$\mathrm{sgn}(d_i)\,\tilde D(\mathbf{x}_i - d_i\,\mathbf{u}_j)$.

As a heuristic, we also defer the initialization for any sample
whose sphere intersects no neighboring sphere, so that the
high-confidence tangent points are committed first. Once these
constraints are folded into the interpolant, the next outer
iteration's $\tilde D$ is more accurate, and the Fibonacci-lattice
search lands on a better warm-start for these deferred samples.
We mark their directions as empty in the current iteration and
re-initialize them by the same lattice search against the updated
$\tilde D$ in the next iteration.

\paragraph{Filtering invalid projections}\label{para:filter-projections}
Projected points that fall outside the exposed region of their
sphere are culled; the alternatives of accepting them as-is or projecting them onto the exposed region boundary
underperform empirically. %
The RT neighbors found for
the short-arc phase are sufficient to filter infeasible projections.
Because each gradient update can
shift the projected point, a sphere whose projection was
feasible in the previous iteration may become infeasible after a
new $\mathbf{g}^{(k+1)}$ is taken. Rather than accept the
regression, we reject any update that turns a previously
feasible projection infeasible and retain the prior $\mathbf{g}$
instead, ensuring the set of feasible projections grows
monotonically across iterations.

\paragraph{Clamping}
\label{sec:clamping}
Sharp edges, like corners, also yield small exposed regions, but
theirs are \emph{skinny} rather than point-like: too elongated for the
short-arc test (Sec.~\ref{sec:short-arc}) to collapse them, yet narrow
enough that the feasibility test above rejects projections that miss
them only slightly. Losing these projections is costly, since the dense
band of tangent points along an edge is what lets the interpolant
reproduce it sharply. We therefore clamp such a projection to the
nearest point on its exposed region boundary after 7 outer iterations,
rather than culling it, 
whenever both its distance to that region and the region's total arc
length fall below $\epsilon_{\mathrm{clamp}} = 0.1$.
Restricting the
clamp to these near-degenerate cases avoids the degradation observed
when boundary projection is applied to every infeasible projection.

\paragraph{De-duplication}\label{sec:deduplication}
Clusters of near-coincident projections may occur, either emitted by
the short-arc step or produced by independent samples' tangent points
converging to the same surface point. This increases per-patch
sample counts and decreases conditioning of the RBF system ~\cite{wendland2004scattered},
without contributing new geometric information.
We de-duplicate points with a fixed radius of $5\times10^{-4}$.

\paragraph{Partition-of-unity weighting}
See Appendix~\ref{sec:acceleration} for details of our partition of unity scheme.

\section{Experiments}
\label{sec:experiments}

In every table, the best score in each column and metric is in bold and
the second best is underlined.

\subsection{Accuracy}
\paragraph{Dataset}
We evaluated our method on a subset of the Thingi10K
dataset~\cite{Zhou2016Thingi10K},
which provides 10,000 3D-printing models curated from Thingiverse.
To ensure well-defined signed distance functions, we restricted the evaluation
set to meshes that are manifold, closed, solid, have no self-intersections,
and contain between 500 and 200,000 vertices. The lower bound excludes
degenerate inputs; the upper bound caps the memory footprint of dense SDF
sampling. From the 5,249 models meeting these criteria, we deterministically
selected the first 300 by ascending file ID. We did not reorient the
meshes to align sharp features with the sampling grid, although most of 
the selected models happen to be modeled axis-aligned.
All models are normalized such that the bounding box is centered at the origin with the longest edge equal to one.
We then sampled its signed distance function on a uniform $\sqrt[3]{n} \times \sqrt[3]{n} \times \sqrt[3]{n}$ grid
within an inflated bounding box (axis-aligned padding of 0.1).
We used libigl \cite{libigl} to compute unsigned distances and generalized winding numbers~\cite{jacobson2013robust} for signs.
We swept $\sqrt[3]{n} \in \{6,$ $10,$ $20,$ $30,$ $40,$ $50,$ $60,$
$80,$ $100\}$ to characterize each method across resolutions.
A gallery of outputs can be seen in Figure~\ref{fig:gallery} and in the supplemental materials.

\paragraph{Baselines and failure modes}
We compared our approach to MES (via the authors' CGAL \emph{Maximal Empty Spheres} module),
RFTA (via the authors' implementation), and Marching Cubes (via scikit-image \cite{scikit-image}).
For RFTA and MES, we evaluated three screening weights $\sigma=1,10,100$.
At very low grid resolutions, the samples may be too sparse for MES to certify any
valid empty spheres, and no contact points are produced. The failure count
per grid size is: $6^3$: 60, $10^3$: 11, $20^3$: 1.
RFTA always produced a mesh at $\sigma=10$, but Poisson Surface Reconstruction occasionally
produces stray isosurface components well beyond the sampled region. We remove 
components with a majority of vertices outside the input samples'
bounding box, falling back to the largest connected component when no 
component satisfies this criterion.
Marching Cubes fails when SDF samples are 
all positive or all negative. The failure count per grid size is: $6^3$: 48, $10^3$: 7, $20^3$: 1.
Our approach succeeded for all inputs.

\begin{figure*}[t]
  \centering
  \includegraphics[width=\linewidth]{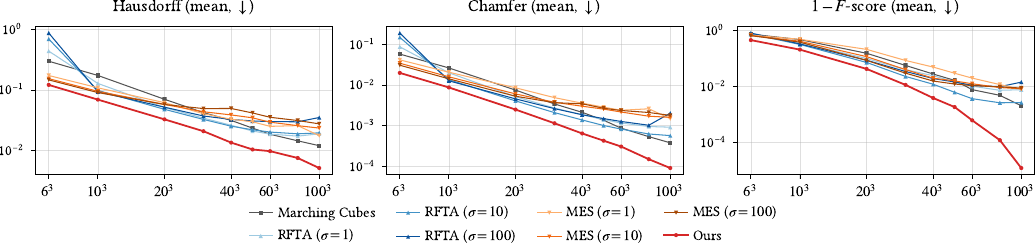}
  \vspace{-1.5em}
  \caption{Accuracy versus number of samples $n$ on 300 Thingi10K
    meshes (mean over successful cases). Lower is better in all
    three panels. RFTA and MES are shown for three screening weights
    $\sigma\!\in\!\{1,10,100\}$. Our method dominates every baseline
    variant at every resolution, and the gap widens with $n$: at
    $n{=}100^3$ the next-best method (MC) has $4.2\times$ our mean
    Chamfer, $2.3\times$ our mean Hausdorff, and more than
    $100\times$ our $1{-}F$ values.}
  \label{fig:accuracy}
\end{figure*}

\paragraph{Metrics}
We report the Hausdorff, Chamfer, and F-score (threshold 1\% of
the unit cube diagonal) on successful cases.
Figure~\ref{fig:accuracy} and Appendix Table~\ref{tab:accuracy-mean}
show that our method achieves strictly lower Hausdorff and Chamfer
error and strictly higher F-score than every baseline variant
across the entire resolution sweep; the gap widens with number of
samples $n$. At $n{=}100^3$
the best RFTA or MES mean Chamfer (RFTA-$\sigma{=}10$,
$5.74{\times}10^{-4}$) is $6.3\times$ ours ($9.15{\times}10^{-5}$),
the best Hausdorff (MES-$\sigma{=}1$) is $3.4\times$ ours
($5.18{\times}10^{-3}$), and our mean F-score is $0.99999$,
with every baseline's error rate $1-F$ more than $100\times$ ours.
At $n{=}10^3$ the best baseline's mean
Chamfer (RFTA-$\sigma{=}100$, $1.27{\times}10^{-2}$) is $1.5\times$
ours ($8.76{\times}10^{-3}$) and F-score is $0.79$ versus $0.68$;
at $n{=}6^3$ the best baseline's Chamfer is again $1.5\times$ ours
($3.09{\times}10^{-2}$ vs.\ $2.00{\times}10^{-2}$, MES-$\sigma{=}100$)
and F-score is $0.54$ versus $0.33$. Our pipeline also returns a mesh on all
$300$ inputs at every resolution tested, while MC and MES
respectively fail on $48$ and $60$ inputs at $n=6^3$.

\paragraph{Ablations}
See the supplemental materials for ablation experiments evaluating partition of unity, short-arc seeding, the choice of RBF kernel, and the tangent-point collapse threshold.

\paragraph{Sharp features}
\label{sec:sharp_features}
To assess sharp-feature reconstruction,
we additionally evaluated the \emph{Edge Chamfer Distance} \cite{chen2021nmc} on the
$50$ meshes with the most sharp features
(using \citet{chen2021nmc}'s sharpness criteria).
Our method had the lowest Edge Chamfer Distance at four of five tested resolutions (Table~\ref{tab:accuracy-mean-sharp}).

\begin{table}
    \revcaption
    \caption{Mean Hausdorff, Chamfer, and edge Chamfer distance \cite{chen2021nmc} over
    the $50$-mesh sharp-feature Thingi10K subset (Sec.~\ref{sec:sharp_features}).}
    \label{tab:accuracy-mean-sharp}
    \centering
    \renewcommand{\arraystretch}{0.9}
    \rowcolors{2}{gray!10}{white}
    \resizebox{\columnwidth}{!}{%
    \begin{tabular}{
        cccc
    }
    \toprule
    \rowcolor{tabheader}
    Method & Hausdorff ($\times 10^{-3}$) & Chamfer
    ($\times 10^{-4}$) & Edge Chamfer ($\times 10^{-4}$) \\
    \midrule
    \rowcolor{tabgroup}
    \multicolumn{4}{c}{$n = 30^3$} \\
    \midrule
    RFTA              & $31.8$ & $31.7$ & $\mathbf{14.6}$ \\
    MES               & $49.1$ & $59.8$ & $26.1$ \\
    MC                & $56.0$ & $49.5$ & $116$ \\
    Ours w/o clamping & \underline{$30.6$} & \underline{$22.6$} & $18.2$ \\
    Ours              & $\mathbf{30.2}$ & $\mathbf{22.0}$ & \underline{$16.4$} \\
    \midrule
    \rowcolor{tabgroup}
    \multicolumn{4}{c}{$n = 60^3$} \\
    \midrule
    RFTA              & $23.0$ & $17.6$ & $13.0$ \\
    MES               & $34.7$ & $38.6$ & $12.6$ \\
    MC                & $26.4$ & $13.7$ & $90.7$ \\
    Ours w/o clamping & $\mathbf{17.8}$ & \underline{$7.26$} & \underline{$11.7$} \\
    Ours              & \underline{$17.9$} & $\mathbf{6.97}$ & $\mathbf{9.42}$ \\
    \midrule
    \rowcolor{tabgroup}
    \multicolumn{4}{c}{$n = 100^3$} \\
    \midrule
    RFTA              & $21.9$ & $14.6$ & $22.6$ \\
    MES               & $44.1$ & $44.5$ & $16.1$ \\
    MC                & $20.1$ & $7.20$ & $34.6$ \\
    Ours w/o clamping & \underline{$5.98$} & \underline{$2.71$} & \underline{$5.51$} \\
    Ours              & $\mathbf{5.92}$ & $\mathbf{2.52}$ & $\mathbf{3.83}$ \\
    \midrule
    \rowcolor{tabgroup}
    \multicolumn{4}{c}{$n = 150^3$} \\
    \midrule
    RFTA              & $31.5$ & $39.8$ & $33.4$ \\
    MES               & $39.4$ & $41.5$ & $78.0$ \\
    MC                & $15.4$ & $3.23$ & $22.3$ \\
    Ours w/o clamping & $\mathbf{3.97}$ & \underline{$1.07$} & \underline{$2.93$} \\
    Ours              & \underline{$4.05$} & $\mathbf{0.968}$ & $\mathbf{2.13}$ \\
    \midrule
    \rowcolor{tabgroup}
    \multicolumn{4}{c}{$n = 200^3$} \\
    \midrule
    RFTA              & $19.8$ & $13.5$ & $26.6$ \\
    MES               & $56.3$ & $64.9$ & $202$ \\
    MC                & $13.1$ & $1.89$ & $13.2$ \\
    Ours w/o clamping & \underline{$3.36$} & \underline{$0.601$} & \underline{$2.16$} \\
    Ours              & $\mathbf{3.29}$ & $\mathbf{0.534}$ & $\mathbf{1.79}$ \\
    \bottomrule
    \end{tabular}
    }
    \vspace{-1em}
\end{table}

We attribute our accuracy to several factors.
Estimated tangent points are hard constraints rather than smoothed as
in the sPSR algorithm used by MES and RFTA.
Unlike MES and RFTA, our RBF interpolation does not explicitly consider normals,
instead relying on off-surface samples.
Indeed, we experimented with Hermite RBF interpolation \cite{macedo2011hermite},
but obtained worse performance since gradients are not well-defined at sharp features.
In MES and RFTA, inaccurate tangent points imply inaccurate normals.

\begin{figure}[t]
  \centering
  \includegraphics[width=.9\linewidth]{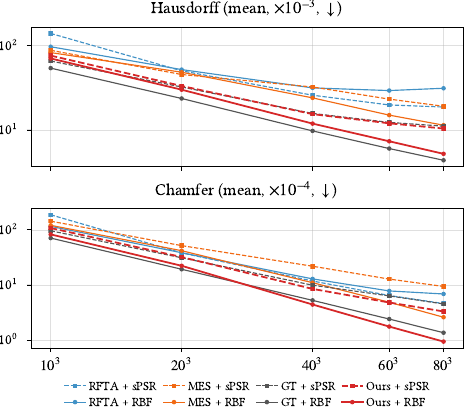}
  \caption{Every combination of tangent-point source (color) and
    reconstruction method (solid with circles: RBF; dashed with squares:
    sPSR) on the first $50$ models of our Thingi10K subset. GT are the 
    ground truth closest surface points for each sample.}
  \label{fig:tp-recon}
\end{figure}

\paragraph{Is our improved performance due to better tangent points or RBF interpolation?}
To understand our algorithm's improved performance,
we evaluated different combinations of tangent points (ours, RFTA, MES, and ground truth) and surface reconstruction methods (sPSR and RBF)
on the first 50 models of our Thingi10K subset.
We used an sPSR screening value of 10 and filtered spurious surfaces from RFTA and MES output.
Results can be seen in Figure~\ref{fig:tp-recon}; the underlying
numbers are in Appendix Tables~\ref{tab:tp-recon-hausdorff}
and~\ref{tab:tp-recon-chamfer}.

For three sources of tangent points (ours, MES, ground truth), RBF reconstruction substantially (Chamfer) or somewhat (Hausdorff) improves error versus sPSR. For RFTA, the effect is reversed; RBF harms error a little (Chamfer) or a lot (Hausdorff).
The Chamfer distance is similar for RBF reconstruction of our tangent points versus ground truth tangent points.
This suggests that, for most methods, RBF reconstruction is superior to sPSR.

For either reconstruction method, our tangent points outperform MES, RFTA, and often ground truth.\footnote{\citet{kohlbrenner2026contouring} also observed suboptimal ground truth performance.}
Thus our method's improved performance can be attributed both to our improved tangent points and the RBF reconstruction.

\begin{table}
    \caption{Robustness to irregular sampling. Mean
    Hausdorff\,/\,Chamfer distance (both $\times 10^{-3}$, $\downarrow$)
    over $10$ shapes with $50^3$ samples placed on a regular
    grid versus at scattered positions.}
    \label{tab:robust-scattered}
    \centering
    \rowcolors{2}{gray!10}{white}
    \resizebox{.5\columnwidth}{!}{%
    \begin{tabular}{
        ccc
    }
    \toprule
    Method & Grid & Scattered \\
    \midrule
    RFTA  & $18.5$\,/\,\underline{$0.850$} & $21.0$\,/\,$0.883$ \\
    MES   & \underline{$16.8$}\,/\,$1.08$ & \underline{$19.2$}\,/\,\underline{$0.837$} \\
    MC    & $26.7$\,/\,$1.36$ & $-$ \\
    Ours  & $\mathbf{10.3}$\,/\,$\mathbf{0.349}$ & $\mathbf{10.2}$\,/\,$\mathbf{0.327}$ \\
    \bottomrule
    \end{tabular}
    }
\end{table}
\paragraph{Scattered data}
We obtain similar Chamfer, Hausdorff, and F1 scores with $n$ points distributed
randomly
rather than arranged in a grid (Table \ref{tab:robust-scattered}).

\paragraph{Truncated values}
When samples farther than an absolute threshold $b$ are deleted, our approach
degrades far more gracefully than RFTA and MES (Table~\ref{tab:robust-truncation}). On the \textsc{Rossignol} model at
$n = 50^3$, our error is essentially unchanged down to $b = 0.05$
($6.80$ versus $6.85$ Hausdorff, $\times 10^{-3}$), and it still recovers a
reasonable surface at $b = 0.002$, whereas at $b = 0.1$ the Chamfer error of
RFTA and MES has already grown $189\times$ and $173\times$.
This may be because the cubic kernel extrapolates well
(Section~\ref{sec:cubic-kernel-extrapolation}).
Marching Cubes clamps out-of-range samples rather than deleting them, so it
retains the sign of the field everywhere and is barely affected by
truncation. It therefore outperforms our approach in Hausdorff distance once
$b \leq 0.005$, although our Chamfer distance stays the lowest at every
bound.

\paragraph{Noise}
We perturbed the distance values with zero-mean Gaussian noise of
standard deviation $\sigma$ (Table~\ref{tab:robust-noise}).
Our method is the most accurate of the four at $\sigma = 0$ (no noise) and $\sigma = 0.001$ in both metrics, and outperforms RFTA and MES at every
noise level tested (up to $\sigma = 0.01$).
However, its advantage over Marching Cubes diminishes as $\sigma$
grows. Marching Cubes attains the lowest Chamfer
distance above $\sigma = 0.005$ and the lowest Hausdorff distance above
$\sigma = 0.01$. This is the cost of our exact interpolation; noise
anywhere in the field enters our reconstruction, while Marching Cubes
reads only cells that straddle the zero crossing.

\begin{table}
    \revcaption
    \caption{Robustness to truncation. Hausdorff\,/\,Chamfer distance
    (both $\times 10^{-3}$, $\downarrow$) on a single model
    (\textsc{Rossignol}) at $n = 50^3$, with the distance field
    truncated to $[-b, b]$. MC clamps to $\pm b$, while the others
    discard out-of-range samples.}
    \label{tab:robust-truncation}
    \centering
    \rowcolors{2}{gray!10}{white}
    \resizebox{\ifdim\width>\columnwidth \columnwidth \else \width \fi}{!}{%
    \begin{tabular}{
        cccccc
    }
    \toprule
    Method & No truncation & $b = 0.1$ & $b = 0.05$ & $b = 0.005$
    & $b = 0.002$ \\
    \midrule
    RFTA  & $17.0$\,/\,\underline{$0.602$} & $491$\,/\,$114$ & $406$\,/\,$69.5$ & $138$\,/\,$16.7$ & $267$\,/\,$30.4$ \\
    MES   & \underline{$13.5$}\,/\,$0.618$ & $510$\,/\,$107$ & $208$\,/\,$38.4$ & $365$\,/\,$57.4$ & $2090$\,/\,$322$ \\
    MC    & $23.0$\,/\,$1.35$ & \underline{$23.0$}\,/\,\underline{$1.35$} & \underline{$23.0$}\,/\,\underline{$1.35$} & $\mathbf{23.0}$\,/\,\underline{$2.17$} & $\mathbf{20.1}$\,/\,\underline{$2.79$} \\
    Ours  & $\mathbf{6.85}$\,/\,$\mathbf{0.280}$ & $\mathbf{6.91}$\,/\,$\mathbf{0.289}$ & $\mathbf{6.80}$\,/\,$\mathbf{0.311}$ & \underline{$24.9$}\,/\,$\mathbf{0.932}$ & \underline{$53.6$}\,/\,$\mathbf{2.67}$ \\
    \bottomrule
    \end{tabular}
    }
\end{table}

\begin{table}
    \revcaption
    \caption{Robustness to noisy distance values. Mean Hausdorff\,/\,Chamfer distance
    (both $\times 10^{-3}$, $\downarrow$) over $10$ shapes at
    $n = 50^3$.
    }
    \label{tab:robust-noise}
    \centering
    \rowcolors{2}{gray!10}{white}
    \resizebox{\ifdim\width>\columnwidth \columnwidth \else \width \fi}{!}{%
    \begin{tabular}{
        ccccc
    }
    \toprule
    Method & $\sigma = 0$ & $\sigma = 0.001$ & $\sigma = 0.005$
    & $\sigma = 0.01$ \\
    \midrule
    RFTA  & $18.5$\,/\,\underline{$0.850$} & $28.1$\,/\,$1.29$ & $44.5$\,/\,$4.74$ & $84.1$\,/\,$11.5$ \\
    MES   & \underline{$16.8$}\,/\,$1.08$ & \underline{$14.8$}\,/\,\underline{$1.11$} & $35.2$\,/\,$4.03$ & $38.7$\,/\,$8.23$ \\
    MC    & $26.7$\,/\,$1.36$ & $26.6$\,/\,$1.52$ & \underline{$24.7$}\,/\,$\mathbf{2.77}$ & $\mathbf{32.2}$\,/\,$\mathbf{4.38}$ \\
    Ours  & $\mathbf{10.3}$\,/\,$\mathbf{0.349}$ & $\mathbf{13.2}$\,/\,$\mathbf{1.08}$ & $\mathbf{21.1}$\,/\,\underline{$3.12$} & \underline{$36.8$}\,/\,\underline{$5.16$} \\
    \bottomrule
    \end{tabular}
    }
\end{table}

\paragraph{High sample counts}
To understand whether our approach outperforms marching cubes at high densities, we ran our method on up to $300^3$ samples on the first 50 meshes in our dataset. For $\sqrt[3]{n} = 40,$ 60, 80, 100, 150, 200, 300, our approach's Chamfer distance is
0.29, 0.31, 0.29, 0.23, 0.31, 0.34, 0.36 $\times$ that of MC, respectively.
Thus our approach produces superior output to MC even at high densities,
whereas MC outperforms RFTA and MES above $n=60^3$ (Figure \ref{fig:accuracy}).

\begin{figure}
  \centering
  \includegraphics[width=\linewidth]{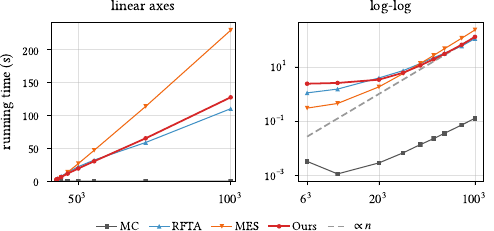}
  \caption{Average runtime (seconds) versus samples $n$ over 30 Thingi10K
    inputs, on linear (left) and logarithmic (right) axes.
    }
  \label{fig:timing}
  \vspace{-1em}
\end{figure}

\subsection{Running Time}
We measured wall-clock time on an Apple MacBook~Pro (Mac17,7) with an M5~Max processor (18 CPU/40 GPU cores) and 128\,GB of unified memory.
RFTA's GPU back-end was enabled.
Figure~\ref{fig:timing} reports the mean over $30$ Thingi10K meshes
against the sample count $n$. Our approach grows almost linearly in $n$. Our approach is faster than MES above $\sqrt[3]{n}\ge 40$; it is $1.8\times$ faster at
$n{=}100^3$.
Our approach is slightly slower than RFTA, by $1.2\times$ at $n{=}100^3$.
Marching Cubes is shown as a runtime lower bound.

\paragraph{Complexity.}
Empirically, regressing the $\log$ of our running time (Fig.~\ref{fig:timing})
against $\log n$ over $\sqrt[3]{n}\ge 30$ gives a slope of ${\sim}0.84$,
so the cost is essentially linear in the input sample count over the tested
range. The slope falls slightly below one because a fixed start-up cost of a
few seconds is amortized as $n$ grows; it rises to ${\sim}0.88$ for
$\sqrt[3]{n}\ge 40$). Combining the per-stage analyses in
Sec.~\ref{sec:short-arc}, Sec.~\ref{sec:iterative-projection}, and
Appendix~\ref{sec:acceleration}, we analyze the
expected cost for well-distributed samples and the worst case.

\emph{Average case.}
Constructing the regular triangulation of $n$ samples takes
expected $O(n \log n)$ time. Its expected vertex degree is
constant, so each representative neighbor set has $O(1)$ size and
the short-arc stage costs $O(n)$ in total.
The iterative projection of Sec.~\ref{sec:iterative-projection}
runs a small constant number of outer iterations.
Each iteration does three things: an inner optimization at every
sample, a feasibility check on every resulting projection, and one
rebuild of the interpolant with the freshly committed tangent points.
The inner optimization takes a bounded number of steps. Each step costs
$O(\log n)$, since a BVH query locates the patches covering the query
point, while bounded patch overlap keeps the blend itself $O(1)$. The
feasibility checks run against constant-size RT neighbor sets. The
rebuild costs $O(n\log n)$ and dominates, so the whole stage is
$O(n\log n)$ and folds into the PU--RBF term above.
The total runtime is:
\[
\underbrace{O(n\log n)}_{\text{RT construction}}
\;+\;
\underbrace{O(n)}_{\text{short-arc}}
\;+\;
\underbrace{O(n\log n)}_{\substack{\text{PU--RBF \,\&}\\ \text{iterative projection}}}
\;=\; O(n\log n)
\]

\emph{Worst case.}
A three-dimensional regular triangulation guarantees no constant
bound on vertex degree: it may contain $O(n^2)$ simplices and take
$O(n^2)$ time to construct, and a single sample may acquire $O(n)$
representative neighbors, in which case the short-arc stage
degenerates to the unrestricted $O(n\,k^2) = O(n^3)$ bound above.
Reaching this bound, however, requires nearly degenerate sample
distributions, e.g., points on two skew lines (with equal
weights)~\cite{erickson2003nice}.

\begin{figure}
  \centering
  \includegraphics[width=\linewidth]{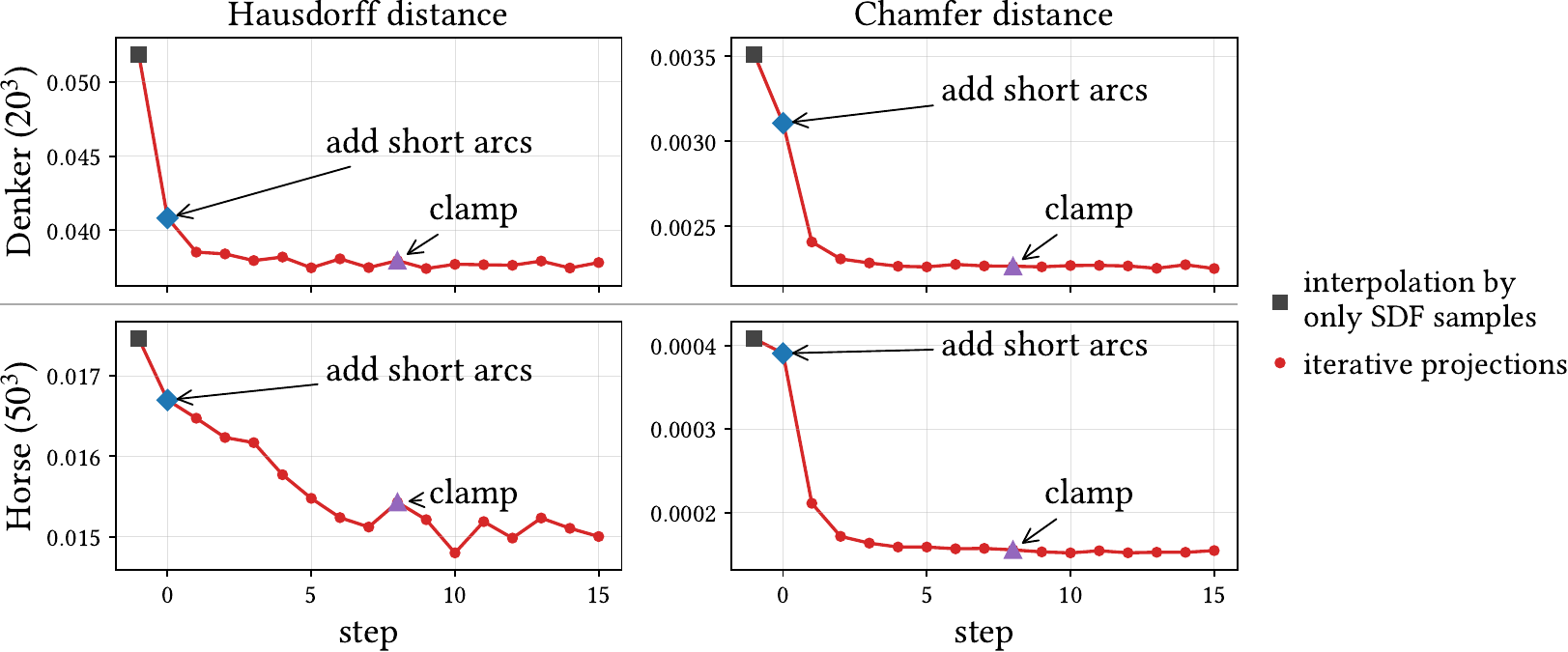}
  \vspace{-2em}
  \caption{Convergence of iterative projection on two representative
    inputs.}
  \label{fig:iter_projection}
\end{figure}

\paragraph{Convergence}
The iterative projection converges within only a handful of steps on
every input we have tested; we always run a constant 10 iterations.
Figure~\ref{fig:iter_projection} shows the per-step error on
two representative inputs. The Chamfer distance drops sharply over
the first few steps and then quickly plateaus, indicating that the
majority of feasible projections are recovered early and further
refinement yields diminishing returns. This justifies our fixed step
budget. The Hausdorff distance follows the same overall trend; the
small upticks visible on \textsc{Armadillo} reflect its sensitivity
to a single worst point, while the Chamfer distance, which averages
over all points, continues to decrease monotonically.

\section{Conclusion \& Limitations}
\label{sec:conclusion}
We presented a radial basis function (RBF)
interpolation of signed distance field samples that pairs every input
sample $(\mathbf{x}_i, d_i)$ with a point
$(\mathbf{y}_i, 0)$ on its implied tangent sphere.
The reconstructed zero level
set passes exactly through the recovered tangent points.
Avoiding a Hermite term keeps the formulation well-defined 
at sharp features, where surface normals can be multi-valued.
We proposed an $S^2$ gradient-based optimization to refine each tangent
point and an analytic boundary construction to handle samples whose
exposed regions collapse.
Our approach scales efficiently with partition of unity.

\begin{figure}
    \centering
    \includegraphics[width=\linewidth]{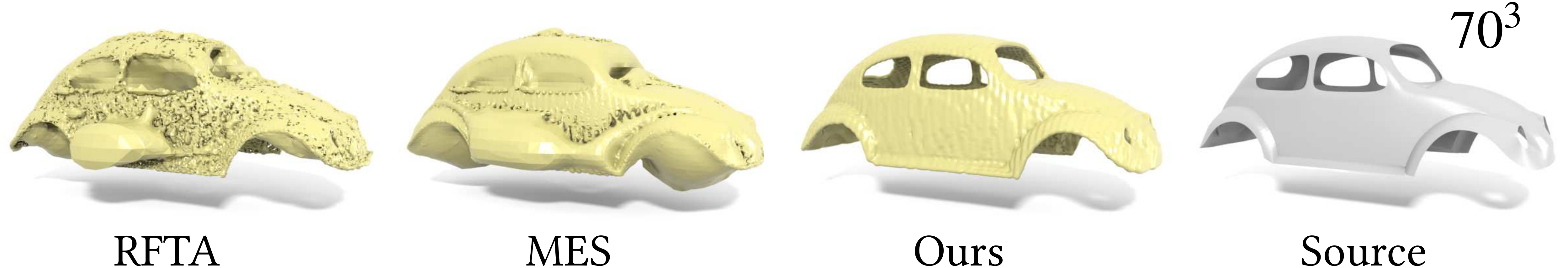}
    \vspace{-2em}
    \caption{Given samples of an unsigned distance field of an open
    surface, our method extracts a thin closed shell around it by
    contouring a small positive isovalue ($0.005$ in this example).}
    \label{fig:open_surfaces}
    \vspace{-1em}
\end{figure}

The interpolant supports evaluation at arbitrary locations
and could drive adaptive sample-insertion schemes such as
RCR~\cite{kohlbrenner2025power}.
Figure~\ref{fig:open_surfaces} shows an example where our
approach is able to extract an open surface
by contouring a small, positive isovalue on UDF.
This is possible because $\tilde D$ approximates a distance field
rather than an indicator function as in sPSR, so a positive isovalue is the offset surface at that distance.
We leave a general exploration of unsigned distance fields as future work.

Several limitations remain.
$\tilde D$ is only an
approximate SDF and is not guaranteed to satisfy the Eikonal
equation $\|\nabla\tilde D\|=1$, so the aforementioned adaptive-insertion schemes
would need care; we leave this direction to future
work.
Furthermore, our kernel is isotropic and so has no preferred
direction along a crease. We reproduce a sharp edge not because the
interpolant represents the crease itself, but because the tangent
points recovered along the edge are dense enough that the residual
undulation between them is imperceptible. Recovering sharp features
from sparser tangent points may call for an anisotropic
formulation \cite{de_marchi_jumping_2020}, which we also leave to future work.
Our extraction resolution is set by a heuristic tied to the input
sample count. We have no criterion certifying that it is fine
enough. A thin structure that $\tilde D$ represents may still be missed
if the extraction grid does not resolve it. The grid can always be
refined further, but choosing a sufficient resolution is currently left
to the user.

\begin{acks}
We are grateful for the anonymous reviewers' constructive suggestions which greatly improved the paper.
This work was supported by the \grantsponsor{usnsf}{United States National Science Foundation}{http://nsf.gov/} (\grantnum{usnsf}{IIS-2402893}).
\end{acks}

\bibliographystyle{ACM-Reference-Format}
\bibliography{references}

\begin{figure*}
    \centering
    \includegraphics[height=8.2in]{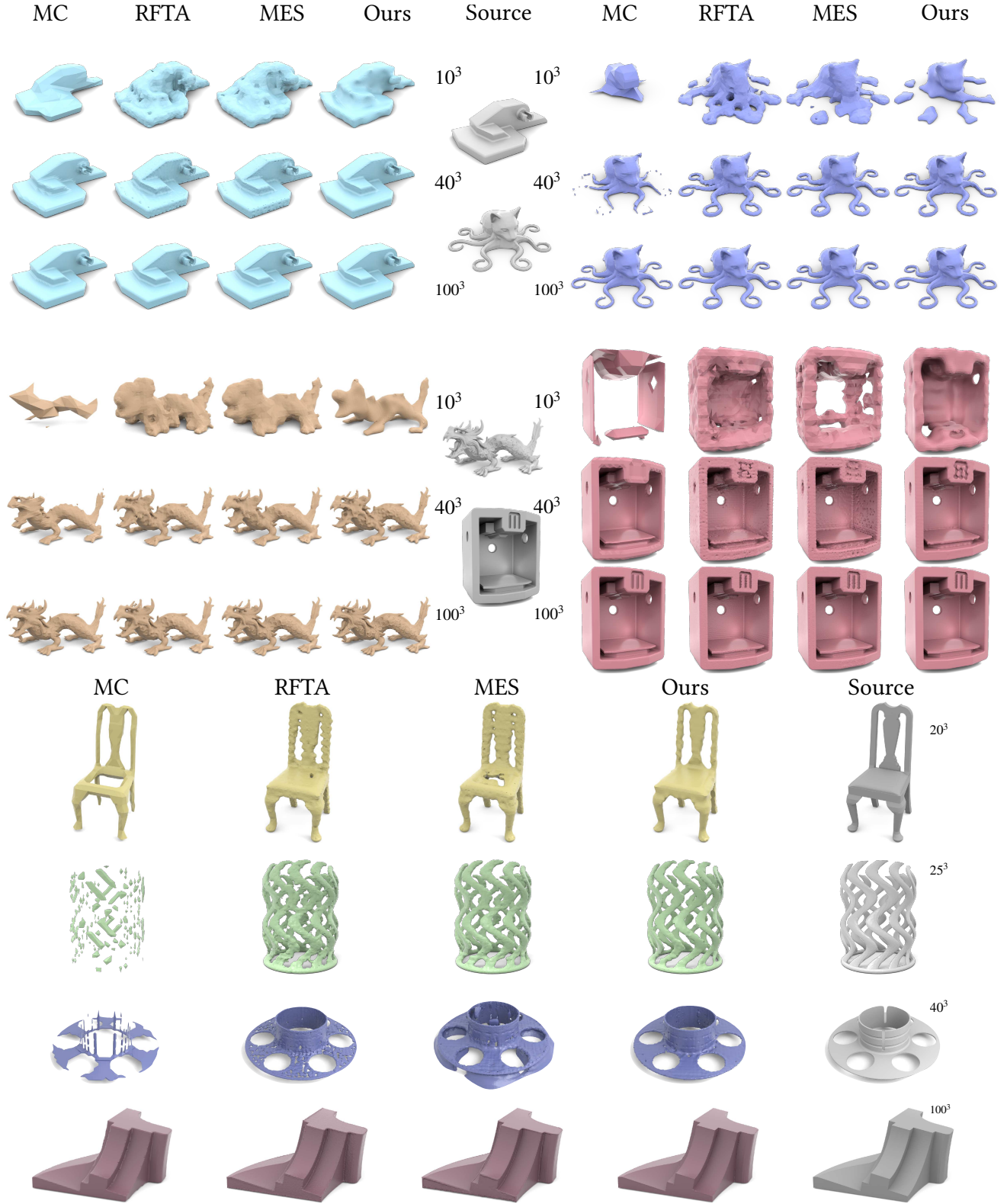}
    \caption{Reconstructions produced by our method, MES, RFTA, and Marching Cubes at varying sampling grid densities.}
    \label{fig:gallery}
\end{figure*}

\clearpage
\appendix

\section{Algorithm Pseudocode}
\label{supp:pseudocode}

This section spells out the pipeline of the main paper's
Section~\ref{sec:algorithm}.
Algorithm~\ref{alg:boundary-arcs} computes the boundary arcs of one
sample sphere's exposed region; Algorithm~\ref{alg:short-arcs} wraps it
into the short-arc tangent-point construction
(Sec.~\ref{sec:short-arc}); and Algorithm~\ref{alg:rbf-sdf} combines
that construction with the iterative projection loop
(Sec.~\ref{sec:iterative-projection}) to produce the final mesh.

Algorithm~\ref{alg:boundary-arcs} processes the neighbors of a
sample sphere $S_i$ one at a time, and the update for each newly added
cap $C_{\mathrm{new}}$ is bidirectional. We write $\mathcal{K}_i$ for
the caps carved on $S_i$ and $\mathcal{B}_i$ for the surviving boundary
arcs, reserving $\mathcal{C}$ for the constraint set of
Algorithm~\ref{alg:rbf-sdf}.

\begin{algorithm}
\caption{\textsc{BoundaryArcs}: exposed-region boundary on a sample sphere $S_i$.}
\label{alg:boundary-arcs}
\begin{algorithmic}[1]
\REQUIRE Caps $\mathcal{K}_i$ carved on $S_i$ by its intersecting
         neighbors (Sec.~\ref{sec:short-arc}).
\ENSURE  Boundary arcs $\mathcal{B}_i = \{a_\ell\}_\ell$ of $S_i$'s exposed region.
\STATE $\mathcal{B}_i \gets \emptyset$;\quad
       $\mathcal{K}_{\mathrm{prev}} \gets \emptyset$
       \COMMENT{Caps processed so far}
\FOR{each cap $C_{\mathrm{new}} \in \mathcal{K}_i$}
    \FOR[Step 1: clip existing arcs]{each arc $a_\ell \in \mathcal{B}_i$}
        \STATE drop the portion of $a_\ell$ inside $C_{\mathrm{new}}$
    \ENDFOR
    \STATE $\mathcal{B}_{\mathrm{new}} \gets$ full circle $[0,2\pi]$ of $C_{\mathrm{new}}$
           \COMMENT{Step 2: new cap's arcs}
    \FOR{each cap $C \in \mathcal{K}_{\mathrm{prev}}$}
        \STATE drop the portion of $\mathcal{B}_{\mathrm{new}}$ inside $C$
    \ENDFOR
    \STATE $\mathcal{B}_i \gets \mathcal{B}_i \cup \mathcal{B}_{\mathrm{new}}$;\quad
           $\mathcal{K}_{\mathrm{prev}} \gets \mathcal{K}_{\mathrm{prev}} \cup \{C_{\mathrm{new}}\}$
\ENDFOR
\RETURN surviving arcs $\mathcal{B}_i$
\end{algorithmic}
\end{algorithm}

\textsc{ShortArcs} (Algorithm~\ref{alg:short-arcs}) declares a sample
sphere degenerate when the total length of its exposed-region boundary
falls below $\epsilon_{\mathrm{degen}}$, and emits the midpoint of each
surviving arc as a candidate tangent point. A sphere with a
non-degenerate exposed region returns $\mathcal{A}_i = \emptyset$ and is
left to the iterative projection loop.

\begin{algorithm}
\caption{\textsc{ShortArcs}: candidate tangent points from collapsed exposed regions.}
\label{alg:short-arcs}
\begin{algorithmic}[1]
\REQUIRE SDF samples $\{(\mathbf{x}_i, d_i)\}_{i=1}^n$;
         collapse threshold $\epsilon_{\mathrm{degen}}$ (default $10^{-5}$).
\ENSURE  Per-sphere candidate tangent points $\{\mathcal{A}_i\}_{i=1}^n$.
\FORALL{$i = 1, \ldots, n$ \textbf{in parallel}}
  \STATE $\mathcal{A}_i \gets \emptyset$
  \STATE $\mathcal{K}_i \gets$ caps carved on $S_i$ by its intersecting neighbors
  \IF[A fully exposed $S_i$ cannot be degenerate]{$\mathcal{K}_i \neq \emptyset$}
     \STATE $\mathcal{B}_i \gets \textsc{BoundaryArcs}(\mathcal{K}_i)$
            \COMMENT{Algorithm~\ref{alg:boundary-arcs}}
     \STATE $L_i \gets \sum_{a_\ell \in \mathcal{B}_i}
             \rho_{c(\ell)}\bigl(t_\ell^{\mathrm{end}} - t_\ell^{\mathrm{start}}\bigr)$
            \COMMENT{Exposed boundary length}
     \IF[Exposed region collapsed to a point]{$L_i < \epsilon_{\mathrm{degen}}$}
        \STATE $\mathcal{A}_i \gets \{\,\textsc{Midpoint}(a_\ell) : a_\ell \in \mathcal{B}_i\,\}$
     \ENDIF
  \ENDIF
\ENDFOR
\RETURN $\{\mathcal{A}_i\}_{i=1}^n$
\end{algorithmic}
\end{algorithm}

In Algorithm~\ref{alg:rbf-sdf}, \textsc{OptimizeGradient} performs the
$S^2$ gradient-based refinement of the tangent direction
$\mathbf{g}_i$ at sample $\mathbf{x}_i$ under the current
$\tilde D$. \textsc{Feasible} tests whether the implied
tangent point $\mathbf{y}_i$ lies in $S_i$'s exposed region using
the cached neighbor list with a BVH fallback
(Sec.~\ref{sec:iterative-projection}).
At most one short-arc candidate per sphere is committed, the one whose
initial RBF value $|\tilde D_0|$ is smallest; the spheres that keep a
candidate form the set $\mathcal{T}$ and are excluded from the
projection loop.

\begin{algorithm}
\caption{RBF-SDF interpolation with implied tangent points.}
\label{alg:rbf-sdf}
\begin{algorithmic}[1]
\REQUIRE SDF samples $\{(\mathbf{x}_i, d_i)\}_{i=1}^n$;
         outer iterations $T$.
\ENSURE  Triangle mesh of the zero level set of an interpolant $\tilde D$
         that fits the input SDF samples and the recovered tangent points.
\STATE $\tilde D_0 \gets \mathrm{RBF}\bigl(\{(\mathbf{x}_i, d_i)\}_{i=1}^n\bigr)$
       \COMMENT{Initial interpolant from SDF samples only}
\STATE $\{\mathcal{A}_i\}_{i=1}^{n} \gets \textsc{ShortArcs}\bigl(\{(\mathbf{x}_i, d_i)\}_{i=1}^n\bigr)$
       \COMMENT{Algorithm~\ref{alg:short-arcs}}
\STATE $\mathbf{a}_i^{\star} \gets
        \arg\min_{\mathbf{a} \in \mathcal{A}_i} |\tilde D_0(\mathbf{a})|$
       \COMMENT{Best candidate on each $S_i$ with $\mathcal{A}_i \neq \emptyset$}
\STATE $\mathcal{T} \gets \bigl\{\, i : \mathcal{A}_i \neq \emptyset,\;
        |\tilde D_0(\mathbf{a}_i^{\star})| < \epsilon_{\mathrm{val}} \,\bigr\}$
       \COMMENT{Spheres with an accepted short-arc point}
\STATE $\mathcal{Y} \gets \bigl\{\, \mathbf{a}_i^{\star} : i \in \mathcal{T} \,\bigr\}$
       \COMMENT{At most one tangent point per sphere (Sec.~\ref{sec:short-arc})}
\STATE $\mathcal{C} \gets \{(\mathbf{x}_i, d_i)\}_{i=1}^n
        \,\cup\, \{(\mathbf{y}, 0) : \mathbf{y} \in \mathcal{Y}\}$
\STATE $\tilde D \gets \mathrm{RBF}(\mathcal{C})$
       \COMMENT{Refit with short-arc tangent points}
\FOR{$t = 1, \ldots, T$}
  \IF{$t = 1$}
    \STATE $\mathcal{I}_t \gets \{\, i \notin \mathcal{T} : S_i \text{ intersects some neighbor sphere}\,\}$
           \COMMENT{Defer no-neighbor samples on the first iteration}
  \ELSE
    \STATE $\mathcal{I}_t \gets \{\, i : \text{sample $i$ has no committed tangent point}\,\}$
  \ENDIF
  \FORALL{$i \in \mathcal{I}_t$ \textbf{in parallel}}
    \STATE $\mathbf{g}_i \gets \textsc{OptimizeGradient}\bigl(\mathbf{x}_i, d_i, \tilde D\bigr)$
           \COMMENT{$S^2$ minimization of $f$, Sec.~\ref{sec:iterative-projection}}
    \STATE $\mathbf{y}_i \gets \mathbf{x}_i - d_i\,\mathbf{g}_i$
           \COMMENT{Implied tangent point on $S_i$}
    \IF[$\mathbf{y}_i$ lies in $S_i$'s exposed region]{$\textsc{Feasible}(\mathbf{y}_i)$}
       \STATE $\mathcal{C} \gets \mathcal{C} \cup \{(\mathbf{y}_i, 0)\}$
    \ENDIF
  \ENDFOR
  \STATE $\tilde D \gets \mathrm{RBF}(\mathcal{C})$
         \COMMENT{Refit interpolant with newly committed tangent points}
\ENDFOR
\RETURN $\textsc{IsosurfaceExtract}(\tilde D, 0)$
       \COMMENT{Dual Contouring}
\end{algorithmic}
\end{algorithm}

\section{RBF Acceleration}
\label{sec:acceleration}
A direct implementation of
the interpolatory RBF solve is $O(n^3)$ due to dense matrix inversion,
which becomes prohibitive for high-resolution SDF grids.
We reduce this to $O(n \log n)$ with a partition of unity
decomposition~\cite{franke_smooth_1982,ohtake2003mpu}.
We perform the decomposition anew for each iteration's RBF solve.
We cover the bounding domain with
overlapping patches $\{U_j\}_{j=1}^J$.
We solve independent local RBF interpolations $f_j$ from the
samples in each $U_j$. The local solutions are blended into a
global function via a normalized partition of unity,
\begin{equation}
  \label{eq:pu-blend}
  \tilde D(\mathbf{x})
  \;=\; \sum_{j=1}^{J} w_j(\mathbf{x})\, f_j(\mathbf{x}),
  \qquad
  w_j(\mathbf{x}) \;:=\;
    \frac{\hat w_j(\mathbf{x})}{\sum_{k} \hat w_k(\mathbf{x})},
\end{equation}
where each $\hat w_j$ is a Wendland $C^2$ compactly supported
function~\cite{wendland2004scattered},
\begin{equation}
  \label{eq:wendland}
  \hat w_j(\mathbf{x})
  \;=\;
  \left( 1 - \tfrac{\|\mathbf{x} - \boldsymbol\xi_j\|}{R_j}
  \right)_{\!\!+}^{\!4}
  \!\!\left( 4\,\tfrac{\|\mathbf{x} - \boldsymbol\xi_j\|}{R_j} + 1
  \right),
\end{equation}
centered at the patch center $\boldsymbol\xi_j$ with support
radius $R_j$, where $(\,\cdot\,)_+ = \max(\,\cdot\,,0)$.

\begin{figure}
    \centering
    \includegraphics[width=\columnwidth]{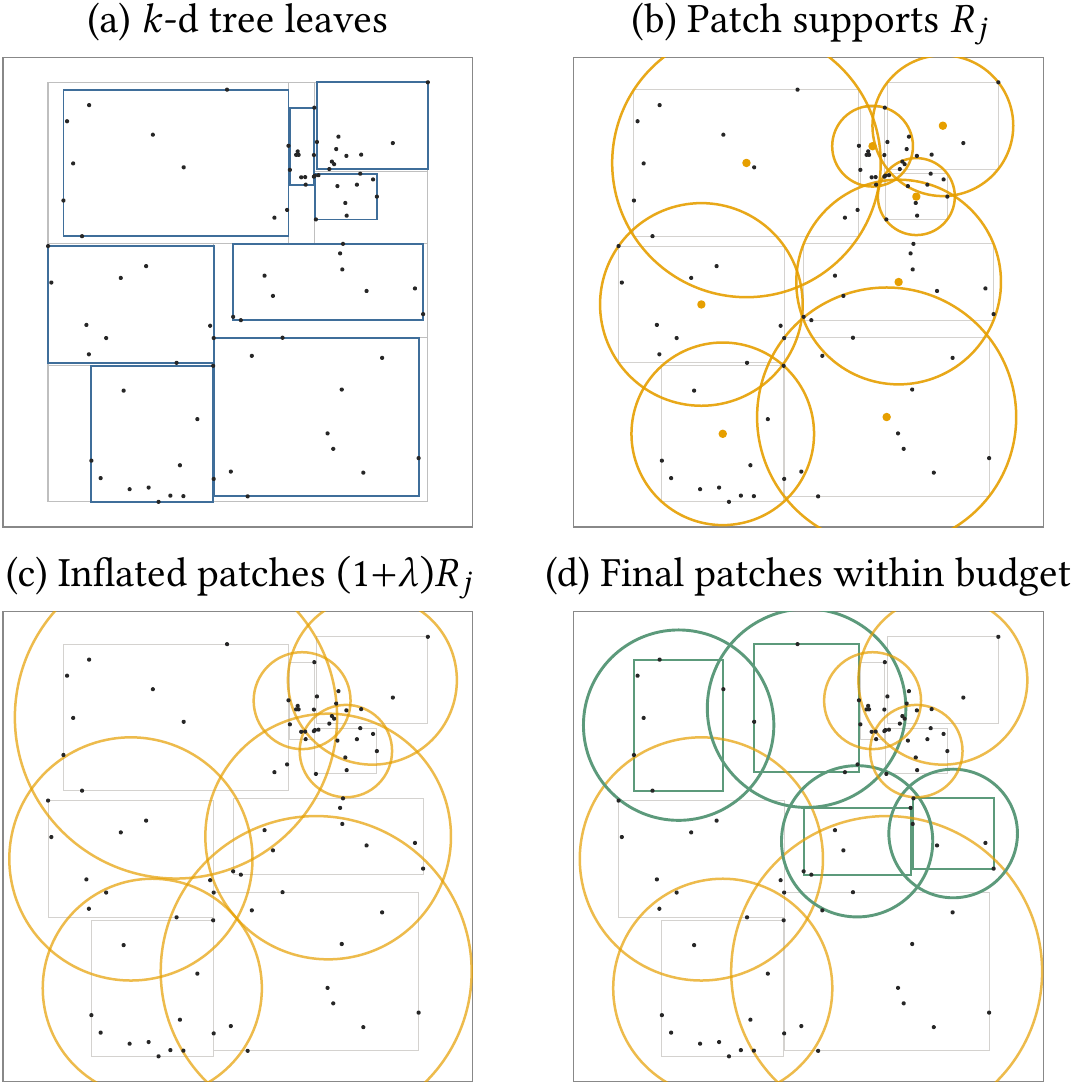}
    \caption{
        Partition-of-unity patch construction.
        (a)~A $k$-d tree partitions the samples into leaves. Grey lines
        show the ambient median-split partition. Blue rectangles
        are each leaf's tight bounding box.
        (b)~Each leaf induces a Wendland support of radius~$R_j$ equal to
        half its bounding-box diagonal.
        (c)~Supports are inflated by $(1{+}\lambda)$ so neighboring
        patches share a non-trivial common region.
        (d)~Patches whose post-inflation sample count exceeds the budget
        are recursively subdivided (green).
    }
    \label{fig:overlap}
\end{figure}

\paragraph{Patch construction.}
We partition the input samples with a $k$-d tree (Fig.~\ref{fig:overlap}a). At each level we
split the current node's bounding box along its longest axis at
the median sample's coordinate,
and recurse until every leaf holds
at most $200$ samples. Each leaf becomes a patch $U_j$, with
center $\boldsymbol\xi_j$ at the leaf's bounding-box center and
support radius $R_j$ equal to half the leaf's bounding-box
diagonal (Fig.~\ref{fig:overlap}b). The Wendland weight $\hat w_j$ is therefore supported on
a ball of radius $R_j$ around $\boldsymbol\xi_j$ that covers the
entire leaf, so every sample in the leaf is included in $U_j$.
We then inflate each patch's support radius 
by a factor of $1+\lambda$ to introduce overlap between neighboring patches (Fig.~\ref{fig:overlap}c).
Inflation can pull in enough additional samples to exceed the
per-patch solve budget, particularly after iterative projection
as projected tangent points cluster
more densely on the surface than the original SDF samples.
Whenever the post-inflation sample
count of a patch exceeds the budget (set to $675$), we recursively
subdivide its leaf before inflation and rebuild the patch (Fig.~\ref{fig:overlap}d). We skip
this subdivision when the leaf already contains only two samples,
to avoid degenerate splits.

We empirically set $\lambda = 0.2$. Larger $\lambda$ increases neighboring-patch
overlap, so adjacent local solves share more samples
and agree more closely on their common region, at the cost of
more patches (and more local solves).

\paragraph{Blended interpolant.}
On each patch $U_j$ we solve the local RBF
system~\eqref{eq:rbf-system} on the subset of samples in $U_j$ to
obtain a local interpolant $f_j$, and blend them via the
partition of unity (Eq.~\ref{eq:pu-blend}). To evaluate at a query
point $\mathbf{x}$, we identify the patches $U_j$ whose supports
contain $\mathbf{x}$, evaluate $f_j(\mathbf{x})$ on each, and
combine them with the normalized weights $w_j(\mathbf{x})$.

In rare cases a query point falls in a small gap between
neighboring patches---outside $\bigcup_j \mathrm{supp}(\hat w_j)$.
We then fall back to the patch of the nearest constraint point
and evaluate its local RBF $f_j$ directly. The cubic
kernel fit to SDF data produces a
smooth interpolant that is approximately Eikonal
($\|\nabla f_j\| \approx 1$) away from the medial axis 
(Figure~\ref{fig:kernels}), so
short-range extrapolation across an inter-patch gap remains
gradient-consistent with the blended interpolant on either
side and introduces no visible discontinuity in $\tilde D$.

A more consequential degeneracy arises when the points within a patch are coplanar or collinear.
(Singular values less than $0.01 \times$ the largest are considered degenerate.)
The local system is then rank-deficient, and the interpolant is unconstrained outside the sample subspace. To address this, we add additional samples to the patch. 
Along the smallest singular value dimension, we add the closest sample to the patch center in both the positive and negative direction that are at least one patch radius away from the subspace. 
The additional samples enter the local interpolation system only; the patch's center, extent, and weight function are unchanged. We repeat this until the system is non-degenerate.

By construction, $w_j(\mathbf{x}_i) > 0$ implies $\mathbf{x}_i$
lies in neighborhood $U_j$, so every active patch at
$\mathbf{x}_i$ interpolates it ($f_j(\mathbf{x}_i) = d_i$). Each
local RBF $f_j$ inherits value interpolation, $C^2$ smoothness,
and linear reproduction from the global RBF. The partition of
unity preserves all three when blending them into $\tilde D$.

\begin{figure}
    \centering
    \includegraphics[width=\columnwidth]{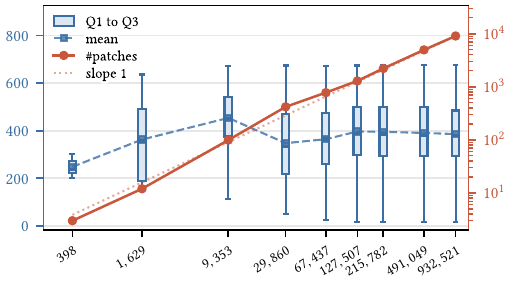}
    \caption{Statistics of partition-of-unity patches across the
    \textsc{chair} test mesh. The boxes plot the distribution of
    per-patch sample counts after subdivision, all below the budget of $675$;
    the solid orange line plots total patch count $J$ vs.\ constraint 
    point count $n$. The dashed orange line plots a reference slope of $O(n)$.
    The $x$-axis (interpolated points) and orange $y$-axis on the right are on a log scale, while the box plot's 
    $y$-axis on the left is linear.
    }
    \label{fig:patch_stats}
\end{figure}

\paragraph{Complexity.}
The decomposition is built with a
$k$-d tree at $O(n \log n)$ cost, followed by bounded re-splitting of
patches that exceed the per-patch budget $m$ after inflation, which
adds at most another $O(n \log n)$ cost. Given this constant
budget, each per-patch solve costs $O(m^3) = O(1)$, and bounded patch
overlap ensures $Jm = O(n)$, where $J$ is the patch count,
so the cumulative solve cost is $O(n)$.
The total cost is therefore dominated by the
$O(n \log n)$ partitioning. %
We empirically verify these assumptions in Fig.~\ref{fig:patch_stats}.
Per-patch sample counts remain concentrated below the budget across our
benchmark, and the total patch count $J$ grows linearly with $n$.
The end-to-end speedup over the global RBF solve is reported in 
Figure~\ref{fig:timing_pu}.

\section{Per-Method Accuracy}
\label{supp:accuracy-details}

\begin{figure*}
    \centering
    \includegraphics[width=\textwidth]{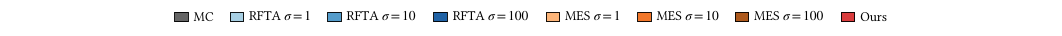}\\[1pt]
    \begin{subfigure}[t]{0.32\textwidth}
        \centering
        \includegraphics[page=1, width=\textwidth]{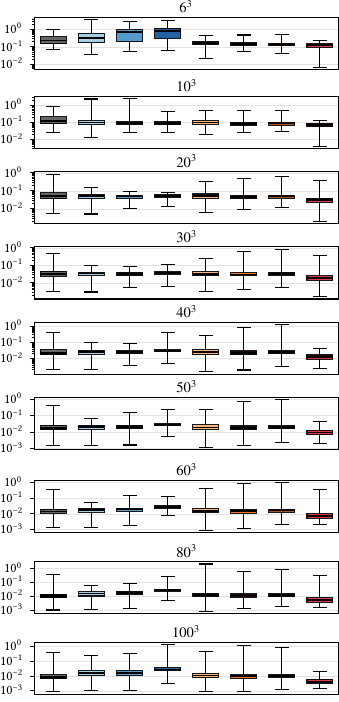}
        \caption{Hausdorff (lower is better, log scale)}
        \label{fig:supp-box-haus}
    \end{subfigure}\hfill
    \begin{subfigure}[t]{0.32\textwidth}
        \centering
        \includegraphics[page=2, width=\textwidth]{accuracy/boxplots.pdf}
        \caption{Chamfer (lower is better, log scale)}
        \label{fig:supp-box-cham}
    \end{subfigure}\hfill
    \begin{subfigure}[t]{0.32\textwidth}
        \centering
        \includegraphics[page=3, width=\textwidth]{accuracy/boxplots.pdf}
        \caption{F1 (higher is better)}
        \label{fig:supp-box-f1}
    \end{subfigure}
    \caption{Per-method, per-resolution distributions of all three
    metrics on the $300$-mesh Thingi10K subset. Each column is one
    metric; each row within a column is one grid resolution, titled
    $n$, from $6^3$ at the top to $100^3$ at the
    bottom. Boxes span Q1--Q3 with a median bar, and whiskers span the
    full range over the meshes (minimum to maximum). The F1 axis is
    fixed to $[0,1]$ on every row so the rows are directly comparable;
    the Chamfer and Hausdorff axes are logarithmic and scaled per row.
    Our method's box
    sits below every baseline's median for Chamfer and Hausdorff
    at every resolution, and dominates F1 most visibly at coarse
    sampling ($\sqrt[3]{n}\!\in\!\{6,10,20\}$).
    }
    \label{fig:supp-boxplots}
\end{figure*}

This section accompanies Figure~\ref{fig:accuracy} of the main
paper with the underlying numerical means
(Table~\ref{tab:accuracy-mean}) and the per-mesh
distributions (Figure~\ref{fig:supp-boxplots}).

\paragraph{Successful-cases denominator.}
Each (method, $\sqrt[3]{n}$) tuple is evaluated on a
$300$-mesh subset of Thingi10K. Every reported metric is
averaged over the cases on which the method produced a non-empty
mesh. Marching Cubes returns no mesh when all SDF samples share
the same sign. This occured $48$, $7$, and $1$ times when
$\sqrt[3]{n}\!\in\!\{6, 10, 20\}$, respectively.
The MES baselines also fail at the same low resolutions
($60$, $11$, and $1$ times, respectively, plus a handful at $\sqrt[3]{n}=100$).
RFTA
and our method both complete on all $300$ inputs at every
tested resolution. The boxplots in
Figure~\ref{fig:supp-boxplots} share the same denominator as the
table. Whiskers omit outliers,
so the long upper
tails of the MES variants at high resolution (visible as
Chamfer/Hausdorff outliers an order of magnitude above their
medians) are reflected in the means but not in the whisker
extents.

\begin{table*}
    \centering
    \caption{Mean F1, Hausdorff, and Chamfer over successful cases
    on the $300$-mesh Thingi10K subset, per (method, $\sqrt[3]{n}$).
    The best score in each column is bolded; ours is best in every column for
    every metric. RFTA and MES variants are shown for screening
    weights $\sigma\in\{1,10,100\}$. The means here are the values
    plotted in Figure~\ref{fig:accuracy} of the main paper.}
    \label{tab:accuracy-mean}
    \setlength{\tabcolsep}{3.5pt}
    \begin{tabular}{lccccccccc}
        \toprule
        Method & \multicolumn{9}{c}{$\sqrt[3]{n}$} \\
        \cmidrule(l){2-10}
        & 6 & 10 & 20 & 30 & 40 & 50 & 60 & 80 & 100 \\
        \midrule
        \multicolumn{10}{l}{\emph{Mean F1 \,($\uparrow$)}} \\
        RFTA $\sigma{=}1$   & 0.279 & 0.629 & 0.898 & 0.959 & 0.976 & 0.985 & 0.990 & 0.993 & 0.992 \\
        RFTA $\sigma{=}10$  & 0.230 & 0.678 & 0.927 & 0.977 & 0.988 & 0.994 & 0.996 & 0.997 & 0.997 \\
        RFTA $\sigma{=}100$ & 0.171 & 0.663 & 0.912 & 0.965 & 0.981 & 0.985 & 0.989 & 0.990 & 0.986 \\
        MES $\sigma{=}1$    & 0.299 & 0.505 & 0.785 & 0.914 & 0.950 & 0.970 & 0.980 & 0.988 & 0.992 \\
        MES $\sigma{=}10$   & 0.327 & 0.579 & 0.882 & 0.959 & 0.980 & 0.984 & 0.987 & 0.991 & 0.992 \\
        MES $\sigma{=}100$  & 0.331 & 0.610 & 0.913 & 0.970 & 0.984 & 0.988 & 0.989 & 0.990 & 0.991 \\
        MC                  & 0.242 & 0.518 & 0.844 & 0.943 & 0.972 & 0.983 & 0.992 & 0.995 & 0.998 \\
        \textbf{Ours}       & \textbf{0.537} & \textbf{0.789} & \textbf{0.957} & \textbf{0.989} & \textbf{0.996} & \textbf{0.998} & \textbf{0.999} & \textbf{1.000} & \textbf{1.000} \\
        \midrule
        \multicolumn{10}{l}{\emph{Mean Hausdorff \,($\downarrow$)}} \\
        RFTA $\sigma{=}1$   & 0.439 & 0.129 & 0.0515 & 0.0344 & 0.0263 & 0.0211 & 0.0188 & 0.0174 & 0.0190 \\
        RFTA $\sigma{=}10$  & 0.694 & 0.0987 & 0.0481 & 0.0328 & 0.0254 & 0.0221 & 0.0203 & 0.0190 & 0.0197 \\
        RFTA $\sigma{=}100$ & 0.871 & 0.0941 & 0.0522 & 0.0374 & 0.0328 & 0.0310 & 0.0302 & 0.0300 & 0.0353 \\
        MES $\sigma{=}1$    & 0.174 & 0.109 & 0.0620 & 0.0408 & 0.0331 & 0.0300 & 0.0252 & 0.0259 & 0.0177 \\
        MES $\sigma{=}10$   & 0.154 & 0.0953 & 0.0580 & 0.0439 & 0.0389 & 0.0349 & 0.0295 & 0.0259 & 0.0236 \\
        MES $\sigma{=}100$  & 0.147 & 0.0909 & 0.0579 & 0.0495 & 0.0497 & 0.0419 & 0.0356 & 0.0314 & 0.0277 \\
        MC                  & 0.299 & 0.174 & 0.0705 & 0.0425 & 0.0316 & 0.0236 & 0.0187 & 0.0146 & 0.0121 \\
        \textbf{Ours}       & \textbf{0.122} & \textbf{0.0696} & \textbf{0.0329} & \textbf{0.0210} & \textbf{0.0136} & \textbf{0.0105} & \textbf{0.00983} & \textbf{0.00758} & \textbf{0.00518} \\
        \midrule
        \multicolumn{10}{l}{\emph{Mean Chamfer \,($\downarrow$)}} \\
        RFTA $\sigma{=}1$   & $8.7{\times}10^{-2}$ & $2.0{\times}10^{-2}$ & $5.2{\times}10^{-3}$ & $2.7{\times}10^{-3}$ & $1.9{\times}10^{-3}$ & $1.4{\times}10^{-3}$ & $1.2{\times}10^{-3}$ & $9.6{\times}10^{-4}$ & $9.3{\times}10^{-4}$ \\
        RFTA $\sigma{=}10$  & $1.5{\times}10^{-1}$ & $1.4{\times}10^{-2}$ & $4.1{\times}10^{-3}$ & $2.1{\times}10^{-3}$ & $1.4{\times}10^{-3}$ & $1.0{\times}10^{-3}$ & $8.3{\times}10^{-4}$ & $6.3{\times}10^{-4}$ & $5.7{\times}10^{-4}$ \\
        RFTA $\sigma{=}100$ & $1.9{\times}10^{-1}$ & $1.3{\times}10^{-2}$ & $4.6{\times}10^{-3}$ & $2.6{\times}10^{-3}$ & $1.9{\times}10^{-3}$ & $1.5{\times}10^{-3}$ & $1.3{\times}10^{-3}$ & $1.0{\times}10^{-3}$ & $2.0{\times}10^{-3}$ \\
        MES $\sigma{=}1$    & $4.2{\times}10^{-2}$ & $2.1{\times}10^{-2}$ & $8.6{\times}10^{-3}$ & $4.9{\times}10^{-3}$ & $3.6{\times}10^{-3}$ & $2.9{\times}10^{-3}$ & $2.4{\times}10^{-3}$ & $2.6{\times}10^{-3}$ & $1.5{\times}10^{-3}$ \\
        MES $\sigma{=}10$   & $3.5{\times}10^{-2}$ & $1.6{\times}10^{-2}$ & $6.1{\times}10^{-3}$ & $3.8{\times}10^{-3}$ & $3.0{\times}10^{-3}$ & $2.6{\times}10^{-3}$ & $2.2{\times}10^{-3}$ & $1.7{\times}10^{-3}$ & $1.6{\times}10^{-3}$ \\
        MES $\sigma{=}100$  & $3.1{\times}10^{-2}$ & $1.4{\times}10^{-2}$ & $5.3{\times}10^{-3}$ & $3.7{\times}10^{-3}$ & $3.5{\times}10^{-3}$ & $2.7{\times}10^{-3}$ & $2.4{\times}10^{-3}$ & $2.1{\times}10^{-3}$ & $1.8{\times}10^{-3}$ \\
        MC                  & $5.8{\times}10^{-2}$ & $2.6{\times}10^{-2}$ & $7.6{\times}10^{-3}$ & $3.4{\times}10^{-3}$ & $2.1{\times}10^{-3}$ & $1.4{\times}10^{-3}$ & $8.8{\times}10^{-4}$ & $5.4{\times}10^{-4}$ & $3.8{\times}10^{-4}$ \\
        \textbf{Ours}       & $\mathbf{2.0{\times}10^{-2}}$ & $\mathbf{8.8{\times}10^{-3}}$ & $\mathbf{2.5{\times}10^{-3}}$ & $\mathbf{1.2{\times}10^{-3}}$ & $\mathbf{6.5{\times}10^{-4}}$ & $\mathbf{4.3{\times}10^{-4}}$ & $\mathbf{3.1{\times}10^{-4}}$ & $\mathbf{1.5{\times}10^{-4}}$ & $\mathbf{9.2{\times}10^{-5}}$ \\
        \bottomrule
    \end{tabular}
\end{table*}

\begin{table*}
    \centering
    \caption{Sensitivity of the algorithm to the total arc length threshold
    $\epsilon_{\mathrm{degen}}$
    on the \textsc{eiffel} mesh at two grid resolutions.
    Candidates are the midpoints of short-arcs.
    \emph{Candidate Quality:} Per-candidate surface distance statistics.
    $\overline{d}_{\mathrm{surf}}$ is the mean. The four $\leq$ columns
    report precision as the fraction of candidates within the listed
    surface-distance threshold.
    The surface-distance measures the distance
    to the ground truth surface.
    \emph{Reconstruction:} The final reconstruction
    quality of the extracted mesh against ground truth.
    Although the number of candidates shrinks with $\epsilon$, up to jitter from parallelism,
    the reconstruction is stable
    across the swept range, as measured by Chamfer and F1.
    Hausdorff fluctuates, as is typical for a worst-case metric among similar outputs.}
    \label{tab:degen-tol}
    \setlength{\tabcolsep}{3pt}
    \begin{tabular}{ccccccccccc}
        \toprule
        & & \multicolumn{6}{c}{Candidate Quality} & \multicolumn{3}{c}{Reconstruction} \\
        \cmidrule(lr){3-8}\cmidrule(lr){9-11}
        $\sqrt[3]{n}$ & $\epsilon_{\mathrm{degen}}$ &
        \#cand. & $\overline{d}_{\mathrm{surf}}$ &
        $\le 10^{-8}$ & $\le 10^{-7}$ & $\le 10^{-6}$ & $\le 10^{-5}$ &
        Hausdorff & Chamfer & F1@$10^{-2}$ \\
        \midrule
        20 & $10^{-4}$ & 2202 & $3.7\!\times\!10^{-8}$ & 99.5\% & 99.5\% & 99.5\% & 100.0\% & 0.0403 & 0.00490 & 0.9052 \\
        20 & $10^{-5}$ & 2181 & $3.1\!\times\!10^{-12}$ & 100.0\% & 100.0\% & 100.0\% & 100.0\% & 0.0339 & 0.00436 & 0.9347 \\
        20 & $10^{-6}$ & 2184 & $3.1\!\times\!10^{-12}$ & 100.0\% & 100.0\% & 100.0\% & 100.0\% & 0.0354 & 0.00454 & 0.9242 \\
        20 & $10^{-7}$ & 2187 & $3.1\!\times\!10^{-12}$ & 100.0\% & 100.0\% & 100.0\% & 100.0\% & 0.0403 & 0.00491 & 0.9077 \\
        20 & $10^{-8}$ & 2120 & $3.2\!\times\!10^{-12}$ & 100.0\% & 100.0\% & 100.0\% & 100.0\% & 0.0359 & 0.00445 & 0.9329 \\
        \midrule
        50 & $10^{-4}$ & 50655 & $6.2\!\times\!10^{-6}$ & 99.0\% & 99.2\% & 99.5\% & 99.6\% & 0.0140 & 0.00099 & 1.0000 \\
        50 & $10^{-5}$ & 50114 & $6.2\!\times\!10^{-6}$ & 99.3\% & 99.4\% & 99.5\% & 99.7\% & 0.0141 & 0.00097 & 1.0000 \\
        50 & $10^{-6}$ & 50055 & $6.2\!\times\!10^{-6}$ & 98.8\% & 98.9\% & 99.2\% & 99.7\% & 0.0140 & 0.00097 & 1.0000 \\
        50 & $10^{-7}$ & 50065 & $6.2\!\times\!10^{-6}$ & 98.6\% & 98.7\% & 99.2\% & 99.7\% & 0.0140 & 0.00097 & 1.0000 \\
        50 & $10^{-8}$ & 48934 & $6.3\!\times\!10^{-6}$ & 98.6\% & 98.8\% & 99.2\% & 99.7\% & 0.0140 & 0.00097 & 1.0000 \\
        \bottomrule
    \end{tabular}

\end{table*}

\section{Ablations}

\subsection{Partition of unity}
\begin{figure*}
  \centering
  \begin{minipage}[b]{0.21\linewidth}
    \centering
    \hfill
    \includegraphics[width=\linewidth]{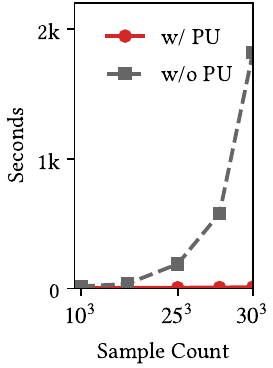}
  \end{minipage}
  \begin{minipage}[b]{0.78\linewidth}
    \centering
    \includegraphics[width=\linewidth]{timing/allMethods.pdf}
  \end{minipage}
  \hfill
  \caption{
    The effect of partition of unity (PU) on the \textsc{Stanford bunny}.
    \textbf{Left:} Without PU, our algorithm takes $1818$ seconds (s)
    at $n=30^3$. With PU, our algorithm takes ranges from $3.45\,$s to $11.04\,$s as $n$ increases from $10^3$ to $30^3$.
    \textbf{Right:} Qualitative comparisons for
    $K=10$ (top) and $K=30$ (bottom).}
  \label{fig:timing_pu}
  \label{fig:qualitative}
\end{figure*}

We measured the wall-clock performance of our algorithm with and without the
partition-of-unity acceleration (Figure~\ref{fig:timing_pu}).
The non-accelerated variant
solves a single global RBF system over all
$n$ samples, which runs in $O(n^3)$ time,
and quickly becomes impractical (already $2169\,$s at
$K=30$). The right panel of Fig.~\ref{fig:qualitative} confirms that
the PU and global reconstructions look very similar at both $K=10$
and $K=30$.

\subsection{Short arc and clamping}
\begin{figure*}
  \centering
  \includegraphics[width=\linewidth]{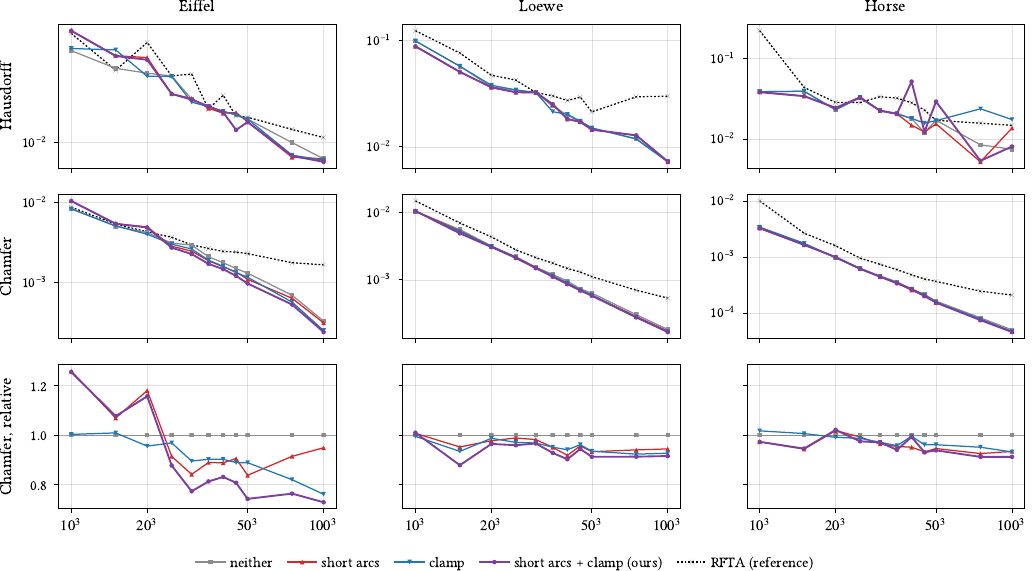}
  \caption{Short-arc seeding and clamping ablated independently.
  The top two rows show absolute Hausdoff and Chamfer distance, respectively, versus RFTA~\cite{sellan2024rfta}. All four variants of our algorithm have lower error than RFTA for most models and resolutions.
  The bottom row shows relative Chamfer distance versus our algorithm without short-arc seeding or clamping.
  At higher sampling rates, both switches improve performance.
  Clamping is neutral on coarse
  samples, whereas short-arc seeding harms a sharp model like \textsc{Eiffel} at low resolution.
  We show Hausdorff in absolute form only; which variant is
  lowest is essentially random.}
  \label{fig:short_arc_ablation}
\end{figure*}

Figure~\ref{fig:short_arc_ablation} evaluates our algorithm with and without
the short-arc seeding (Sec.~\ref{sec:short-arc}) and clamping (Sec.~\ref{sec:clamping})
on three inputs (\textsc{Eiffel}, \textsc{Loewe}, \textsc{horse}).
All four variants of our algorithm, including the simplest version without short-arc seeding or clamping, outperform RFTA~\cite{sellan2024rfta}
for most models and resolutions.
When focusing on the relative Chamfer distance of our variants versus
the simplest version, we see that both switches improve performance
at higher sampling rates.
At low resolutions, clamping is neutral and short-arc seeding harms the
\textsc{Eiffel} example. We have not tested this extensively, but believe
it is due to \textsc{Eiffel}'s sharp features.
Enabling both is best for $n \geq 25^3$ at every resolution on \textsc{Eiffel} and
\textsc{Loewe}, and at all but two on \textsc{Horse},
reaching $0.73\times$ on \textsc{Eiffel} at $n{=}100^3$.
On the smoother \textsc{Loewe} and \textsc{Horse} examples,
the variants have a smaller effect;
the two together save about $8\%$ at the finest sampling.

\begin{figure}
    \renewcommand{\thesubfigure}{\roman{subfigure}}
    \centering
    \begin{subfigure}[b]{0.22\columnwidth}
        \centering
        \includegraphics[width=\linewidth]{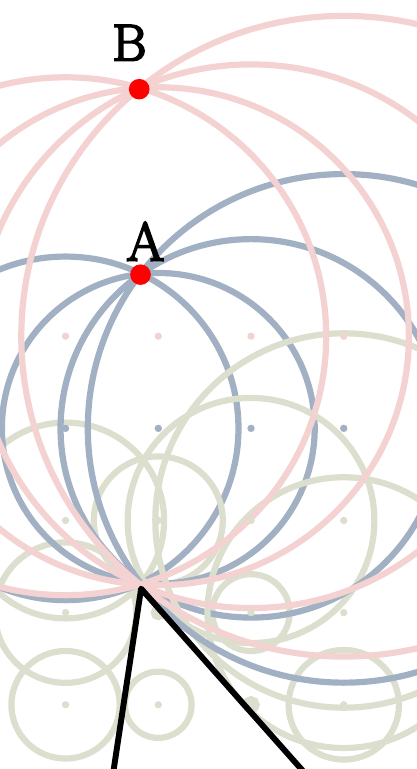}
        \caption{}
        \label{fig:phantom-arc}
    \end{subfigure}\hspace{2pt}%
    \begin{minipage}[b]{0.76\columnwidth}
        \centering
        \begin{subfigure}[b]{\linewidth}
            \centering
            \includegraphics[width=\linewidth]{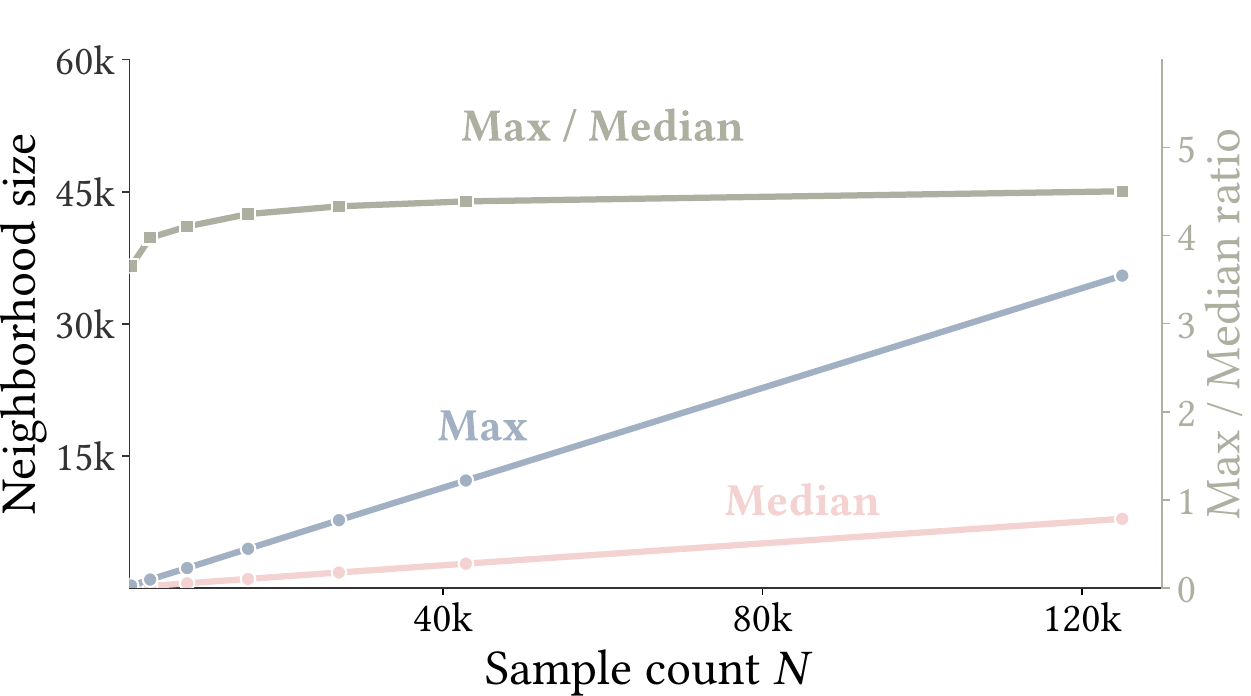}
            \caption{}
            \label{fig:k-vs-n}
        \end{subfigure}\\[4pt]
    \end{minipage}

    \caption{(\subref{fig:phantom-arc}) Phantom points illustrated in
    2D. When the centers are collinear, circles passing through a
    shared corner produce a short arc at the true corner and a mirror
    short arc across the line of centers.
    The interior mirror (A) is clipped by surrounding circles; the
    mirror at the sampling boundary (B) escapes clipping and emits a
    spurious tangent-point candidate.
    (\subref{fig:k-vs-n}) Median and maximum number of intersecting
    neighbors for each sample vs.\ sample count $n$ across the \textsc{bunny} test mesh. Both
    grow linearly with $n$, so the unbounded per-sphere cost scales
    as $O(n^2)$ and the total cost as $O(n^3)$.
    }
    \label{fig:phantom-and-k}
\end{figure}

Figure~\ref{fig:phantom-arc} illustrates a subtler degeneracy. When a
chain of sample spheres shares a corner tangent point and their centers
are nearly collinear, symmetry places a short arc not only at the true
corner but also at its mirror image across the line of centers. In the
interior of the sampling, surrounding spheres clip the mirror arc away;
near a boundary of the sampled region no sphere is available to clip it,
and the mirror arc would otherwise emit a phantom tangent-point
candidate far from the true surface. Disconnected short-arc regions can
be constructed deliberately as well, which is why we screen candidates
by their initial RBF value rather than accepting all of them.

\begin{figure*}
  \centering
    \includegraphics[width=.85\columnwidth]{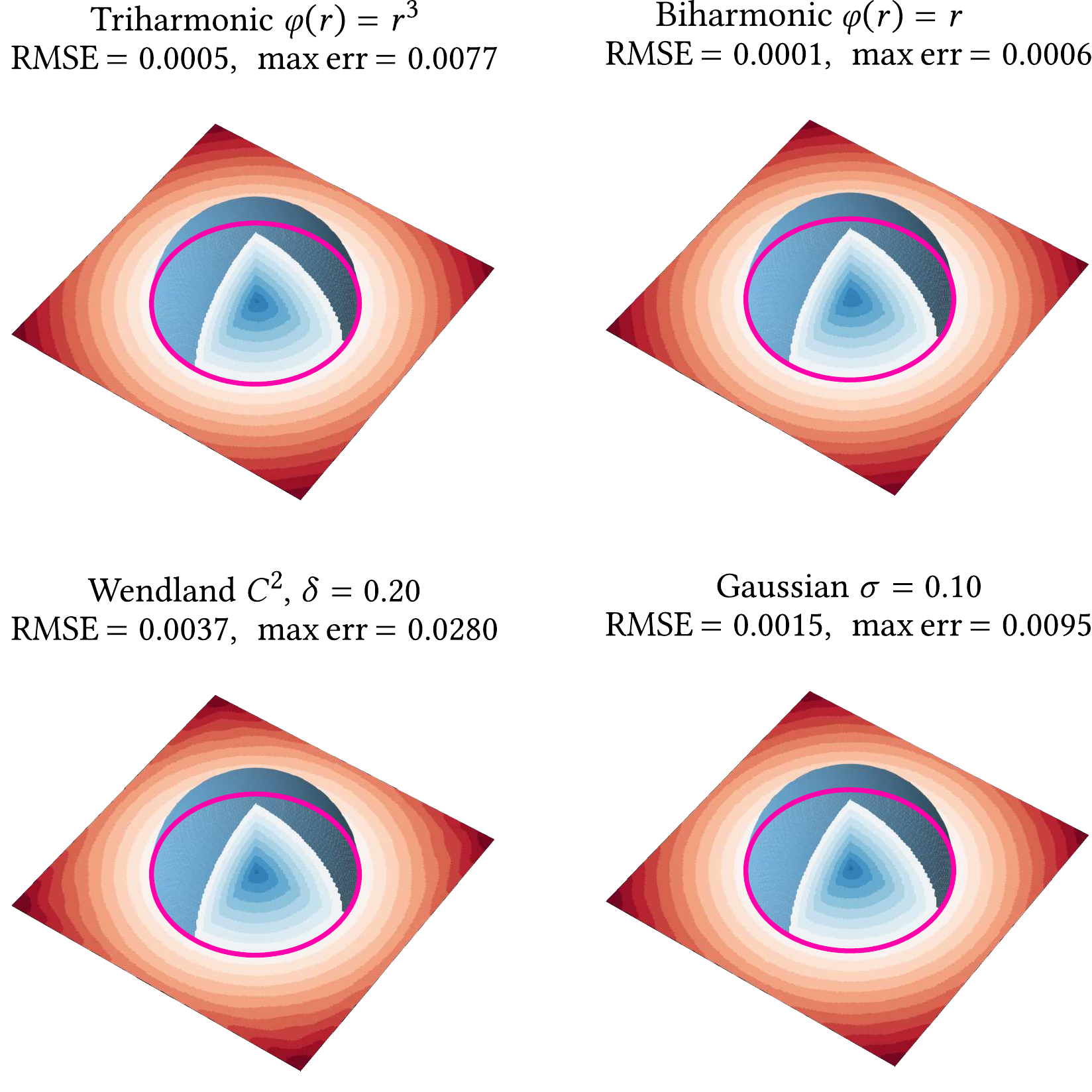}\hfill
  \includegraphics[width=.85\columnwidth]{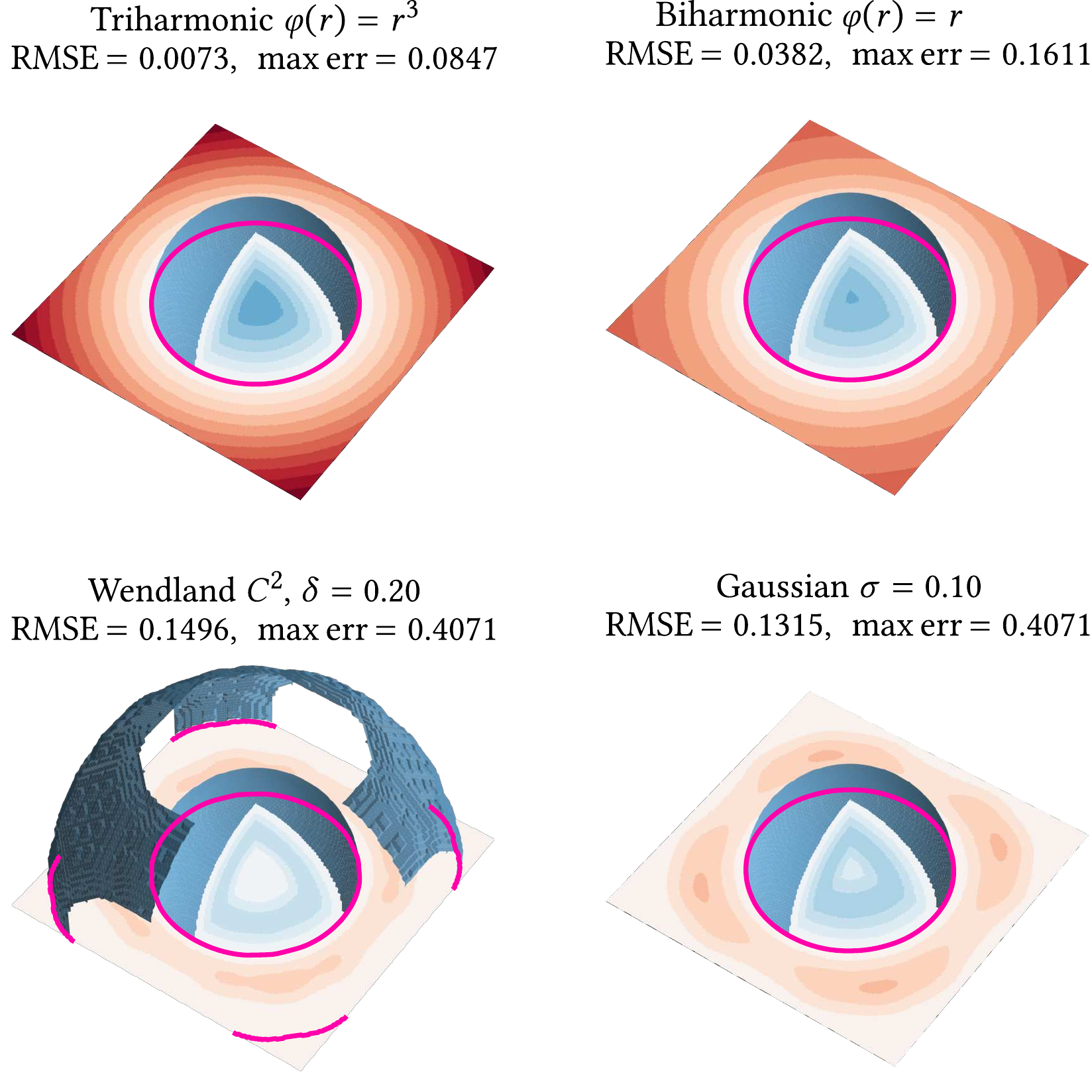}\makebox[1.3cm]{}\\

    \includegraphics[width=\columnwidth]{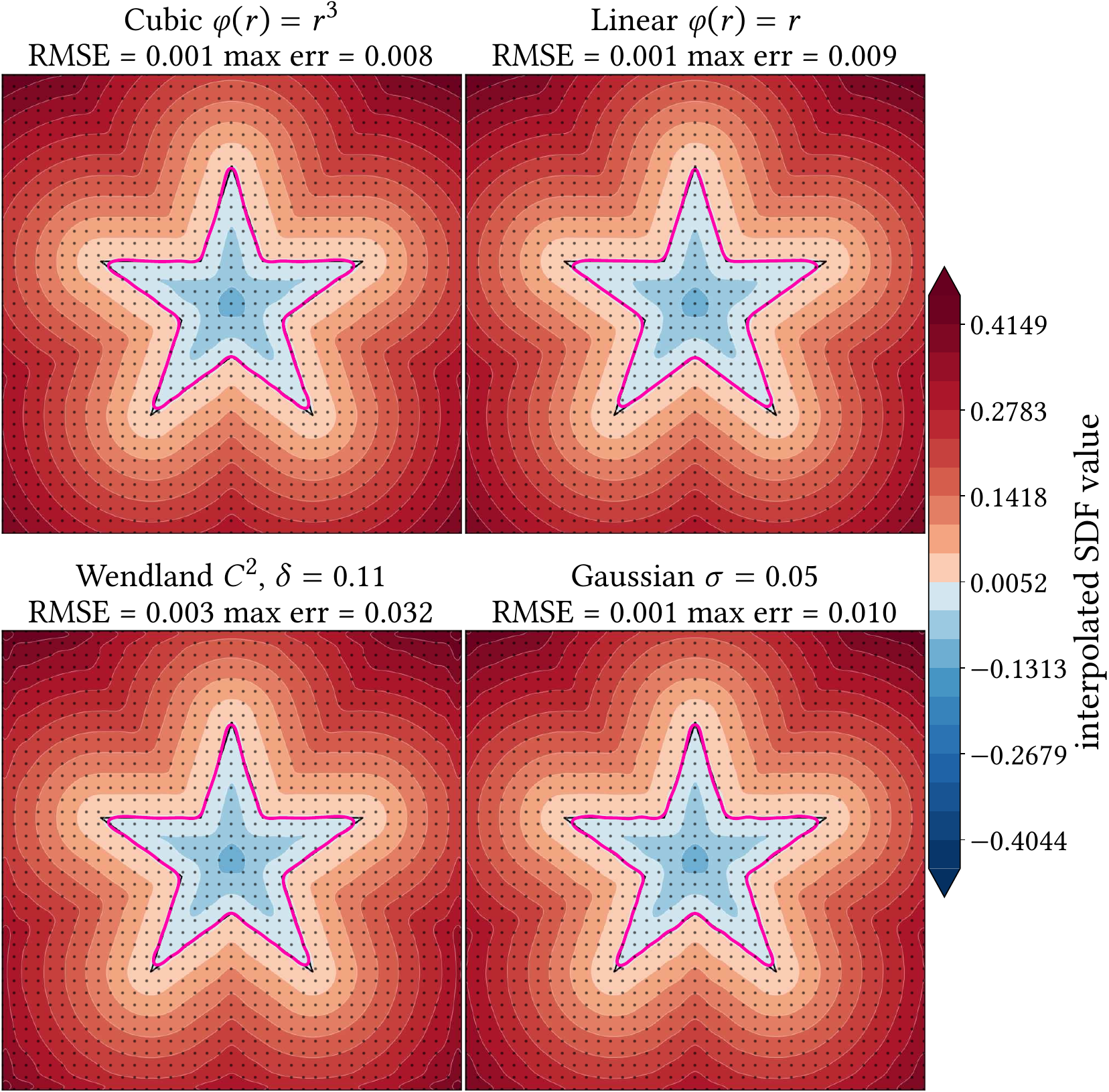}\hfill
  \includegraphics[width=\columnwidth]{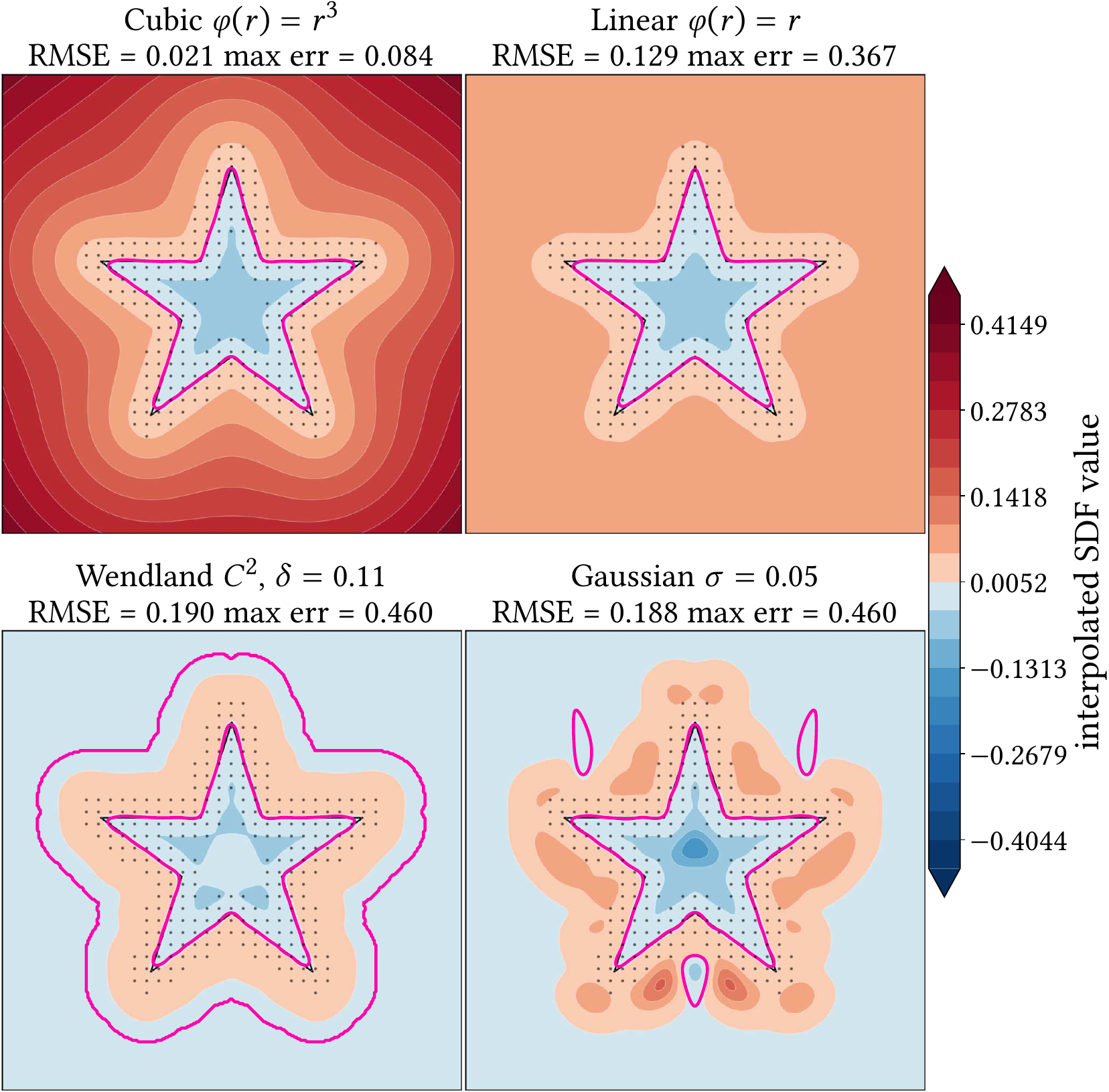}

  \caption{Heatmaps of the reconstructed implicit field
    under four RBF kernels on a sphere in 3D and a star in 2D:
    3D triharmonic ($r^3$, ours), 3D biharmonic ($r$), Gaussian, and Wendland's compactly supported
    $C^2$. Left: With dense samples throughout the domain,
    all kernels have low error. Right: With only a narrow band of samples, only the cubic kernel
    yields a smooth field whose far-field gradient stays closest to unit norm.
    In 2D, we also evaluated the triharmonic ($r^4 \log r$) and biharmonic ($r^2 \log r$) kernels;
    they produced higher error than the cubic kernel.
    }
    \label{fig:kernels}
\end{figure*}
\subsection{RBF kernel}
We compared the cubic (triharmonic in 3D) kernel $\phi(r) = r^3$ used in
our method against four alternatives: $\phi(r) = r$ (biharmonic in 3D),
$\phi(r) = r^2 \log r$ (biharmonic in 3D), the Gaussian kernel, and Wendland's
compactly supported $C^2$ kernel.
Figure~\ref{fig:kernels} shows heatmaps of the
reconstructed implicit field on a representative input under each
kernel. The cubic kernel produces the most stable extrapolation,
with a far-field gradient close to the unit-norm property
$\|\nabla D\| = 1$ of a true SDF.

\subsection{Tangent-Point Collapse Threshold}
\label{supp:degen-tol}

Section~\ref{sec:short-arc} of the main paper introduces a length threshold
$\epsilon_{\mathrm{degen}}$ that decides when the boundary of a sphere's
exposed region has shrunk enough to collapse to a single tangent point. We
use $\epsilon_{\mathrm{degen}} = 10^{-8}$ on meshes normalized to a unit
bounding box. To verify that this choice is reasonable and to characterize
the algorithm's sensitivity to it, we swept
$\epsilon_{\mathrm{degen}} \in \{10^{-4}, 10^{-5}, 10^{-6}, 10^{-7}, 10^{-8}\}$
on the \textsc{eiffel} mesh at two sampling densities (Tab.~\ref{tab:degen-tol}).

Our choice of $\epsilon_{\mathrm{degen}} = 10^{-8}$
is the strictest value in our sweep. It produces tangent-point
candidates closest to the underlying surface (smallest mean surface
distance and highest fraction of candidates within $10^{-5}$ on the test
mesh). The data in Tab.~\ref{tab:degen-tol} serves
as a sensitivity analysis on a single mesh.
It shows
that the algorithm is insensitive to
$\epsilon_{\mathrm{degen}}$ in the tested range.
Chamfer and F1 are similar across this range.
Hausdorff exhibits the minor variability expected of a worst-case distance
on essentially similar input.

\section{Tangent points versus reconstruction}

Data for every combination of tangent-point source and reconstruction
method (Tables~\ref{tab:tp-recon-hausdorff}
and~\ref{tab:tp-recon-chamfer}), plotted in Figure~\ref{fig:tp-recon}
of the main paper.

\begin{table}
    \revcaption
    \caption{Mean Hausdorff distance ($\times 10^{-3}$, $\downarrow$) over
    successful cases for every combination of tangent-point source
    and reconstruction method, on the first $50$ models of our
    Thingi10K subset. GT are the ground truth closest surface points
    for each sample.}
    \label{tab:tp-recon-hausdorff}
    \centering

    \renewcommand{\arraystretch}{0.9}
    \rowcolors{3}{gray!10}{white}
    \resizebox{\columnwidth}{!}{%
    \setlength{\tabcolsep}{4pt}%
    \begin{tabular}{
      cc@{\hspace{4.5pt}}cc@{\hspace{4.5pt}}cc@{\hspace{4.5pt}}cc@{\hspace{4.5pt}}cc@{\hspace{4.5pt}}c
    }
    \toprule
    \rowcolor{white}
     & \multicolumn{2}{c}{$\sqrt[3]{n}{=}10$} & \multicolumn{2}{c}{$\sqrt[3]{n}{=}20$}
     & \multicolumn{2}{c}{$\sqrt[3]{n}{=}40$} & \multicolumn{2}{c}{$\sqrt[3]{n}{=}60$}
     & \multicolumn{2}{c}{$\sqrt[3]{n}{=}80$} \\
    \cmidrule(lr){2-3} \cmidrule(lr){4-5} \cmidrule(lr){6-7} \cmidrule(lr){8-9} \cmidrule(lr){10-11}
    \rowcolor{white}
    Tangent points & sPSR & RBF & sPSR & RBF & sPSR & RBF & sPSR & RBF & sPSR & RBF \\
    \midrule
    RFTA & $139$ & $96.8$ & $50.2$ & $51.9$ & $25.8$ & $31.5$ & $19.8$ & $29.3$ & $18.7$ & $31.2$ \\
    MES  & $87.8$ & $82.5$ & $45.4$ & $48.3$ & $32.3$ & $24.0$ & $23.2$ & $15.0$ & $19.1$ & $11.4$ \\
    GT   & $\mathbf{65.5}$ & $\mathbf{54.0}$ & $\mathbf{32.3}$ & $\mathbf{23.6}$
         & \underline{$15.7$} & $\mathbf{9.75}$ & \underline{$12.3$} & $\mathbf{6.03}$
         & \underline{$11.1$} & $\mathbf{4.38}$ \\
    Ours & \underline{$76.0$} & \underline{$70.3$} & \underline{$33.2$} & \underline{$30.0$}
         & $\mathbf{15.4}$ & \underline{$11.9$} & $\mathbf{12.0}$ & \underline{$7.35$}
         & $\mathbf{10.4}$ & \underline{$5.22$} \\
    \bottomrule
    \end{tabular}
    }
\end{table}

\begin{table}
    \revcaption
    \caption{Mean Chamfer distance ($\times 10^{-3}$, $\downarrow$) over
    successful cases for the same combinations and inputs as
    Table~\ref{tab:tp-recon-hausdorff}.}
    \label{tab:tp-recon-chamfer}
    \centering
    
    \renewcommand{\arraystretch}{0.9}
    \rowcolors{3}{gray!10}{white}
    \resizebox{\columnwidth}{!}{%
    \begin{tabular}{
      cc@{\hspace{4.5pt}}cc@{\hspace{4.5pt}}cc@{\hspace{4.5pt}}cc@{\hspace{4.5pt}}cc@{\hspace{4.5pt}}c
    }
    \toprule
    \rowcolor{white}
     & \multicolumn{2}{c}{$\sqrt[3]{n}{=}10$} & \multicolumn{2}{c}{$\sqrt[3]{n}{=}20$}
     & \multicolumn{2}{c}{$\sqrt[3]{n}{=}40$} & \multicolumn{2}{c}{$\sqrt[3]{n}{=}60$}
     & \multicolumn{2}{c}{$\sqrt[3]{n}{=}80$} \\
    \cmidrule(lr){2-3} \cmidrule(lr){4-5} \cmidrule(lr){6-7} \cmidrule(lr){8-9} \cmidrule(lr){10-11}
    \rowcolor{white}
    Tangent points & sPSR & RBF & sPSR & RBF & sPSR & RBF & sPSR & RBF & sPSR & RBF \\
    \midrule
    RFTA & $18.8$ & $11.6$ & $3.83$ & $3.85$ & $1.18$ & $1.29$ & $0.645$ & $0.780$ & $0.460$ & $0.687$ \\
    MES  & $14.4$ & $12.1$ & $5.18$ & $4.19$ & $2.19$ & $1.10$ & $1.28$  & $0.486$ & $0.941$ & $0.260$ \\
    GT   & $\mathbf{9.63}$ & $\mathbf{7.11}$ & $\mathbf{3.12}$ & $\mathbf{1.93}$
         & \underline{$0.991$} & \underline{$0.526$} & \underline{$0.634$} & \underline{$0.241$}
         & \underline{$0.453$} & \underline{$0.136$} \\
    Ours & \underline{$10.8$} & \underline{$8.29$} & \underline{$3.22$} & \underline{$2.23$}
         & $\mathbf{0.849}$ & $\mathbf{0.442}$ & $\mathbf{0.481}$ & $\mathbf{0.176}$
         & $\mathbf{0.331}$ & $\mathbf{0.0939}$ \\
    \bottomrule
    \end{tabular}
    }
\end{table}

\clearpage

\end{document}